\documentclass[openacc]{rsproca_new}
\usepackage{color}
\usepackage{amsmath,amssymb}
\usepackage[utf8]{inputenc}

\renewcommand{\vec}{\mathbf}
\newcommand{\curl}{\boldsymbol{\nabla}\times}
\newcommand{\bcdot}{\boldsymbol{\cdot}}
\newcommand{\bnabla}{\boldsymbol{\nabla}}
\newcommand{\rot}{\boldsymbol{\Omega}}
\newcommand{\BB}{\vec{B}}
\newcommand{\B}{\vec{b}}
\newcommand{\UU}{\vec{U}}
\newcommand{\U}{\vec{u}}
\newcommand{\e}{\vec{e}}
\newcommand{\Le}{\mathrm{Le}}
\newcommand{\Lu}{\mathrm{Lu}}
\newcommand{\Ek}{\mathrm{Ek}}
\newcommand{\advection}[2]{\left(#1\boldsymbol{\cdot\nabla}\right)#2}

\titlehead{Research}

\begin{document}

\title{Interannual Magneto-Coriolis modes and their sensitivity on the magnetic field within the Earth's core}

\author{F. Gerick$^{1,2}$ and P.W. Livermore$^{3}$}

\address{$^1$ National Centre for Space Studies, Toulouse, France\\
$^2$ Royal Observatory of Belgium, Brussels, Belgium\\
$^3$ School of Earth and Environment, University of Leeds, Leeds, United Kingdom}

\subject{geophysics, fluid mechanics}


\corres{P.W. Livermore\\
\email{P.W.Livermore@leeds.ac.uk}}

\begin{abstract}
Linear modes for which the Coriolis acceleration is almost entirely in balance with the Lorentz force are called Magneto-Coriolis (MC) modes.
These MC modes are thought to exist in Earth's liquid outer core and could therefore contribute to the variations observed in Earth's magnetic field.
The background state on which these waves ride is assumed here to be static and defined by a prescribed magnetic field and zero flow. We introduce a new computational tool to efficiently compute solutions to the related eigenvalue problem, and study the effect of a range of both axisymmetric and non-axisymmetric background magnetic fields on the MC modes.
We focus on a hierarchy of conditions that sequentially partition the numerous computed modes into those which are (i) in principle observable, (ii) those which match a proxy for interannual geomagnetic signal over 1999-2023, and (iii) those which align with core-flows based on recent geomagnetic data. 
We found that the background field plays a crucial role in determining the structure of the modes. In particular, we found no examples of axisymmetric background fields that support modes consistent with recent geomagnetic changes, but that some non-axisymmetric background fields do support geomagnetically consistent modes. 

\end{abstract}

\begin{fmtext}

\end{fmtext}
\maketitle

\section{Introduction}\label{sec:intro}

Earth's magnetic field varies on a broad range of timescales, from billions of years to milliseconds.
Changes in the magnetic field occurring at periods longer than about a year are termed secular variation (SV).
The SV has its origin in flows within the liquid outer core, where Earth's magnetic field is generated by a self-sustaining dynamo action.
Observations of the SV therefore have a direct link to the flow field within the Earth's core, allowing us to better constrain this inaccessible region of our planet.

High resolution maps of SV have been made possible over the last 24~years because of satellite data, providing accurate measurements with global spatial coverage \cite{lesur_rapid_2022}. 
Based on several years of satellite measurements, global models of the internal geomagnetic field such as CHAOS \cite{olsen_chaos_2006, finlay_chaos7_2020} and GRIMM \cite{lesur_grimm_2008, lesur_second_2010} have revealed interannual SV at a global scale across the surface of the Earth \cite{olsen_rapidly_2008}.
We use here and in the rest of the article the word interannual to mean variation over 1~year or more, including up to the 24~years of focus in this work.
The emerging picture of SV is one of decadal to centennial change on the core's convective timescale of about 150 years, superimposed by relatively fast, periodic interannual signals, which strongly suggest a wave origin \cite{chulliat_geomagnetic_2014a, finlay_recent_2016, kloss_timedependent_2019}.
The rapid SV has its largest amplitudes in regions of low latitudes, i.e. close to the geographic equator, and close to the north pole \cite{chulliat_core_2010, chulliat_geomagnetic_2014a, gillet_dynamical_2022}.
A more detailed presentation of the interannual SV observed by satellite data is given in Section \ref{sec:data}.

Variations in Earth's magnetic field have long been postulated to be partially accounted for by global hydromagnetic modes or localized travelling waves within the liquid outer core of the Earth, assumed to be rotating, electrically conducting, inviscid, and incompressible\cite{hide_free_1966,braginsky_magnetic_1967}.
There are three main classes of mode that can exist in the liquid core: inertial, torsional and Magneto-Coriolis, whose dynamics and observational signature we summarise below. 
Other types of modes can exist when making different structural assumptions; for example, a layer of stable stratification supports Magneto-Archimedes-Coriolis waves owing to the additional restoring force through buoyancy \cite{buffett_geomagnetic_2014}. 
We refer the reader to \cite{finlay_course_2008} for a thorough introduction into the topic.

The shortest period modes of the three classes are the inertial modes at near-diurnal periods. 
They exist also in the purely hydrodynamic case and are only slightly modified by electrical conductivity and the presence of a magnetic field. 
Inertial modes, which include the quasi-geostrophic inertial modes (related to Rossby modes) \cite{zhang_inertial_2001}, have been studied extensively in laboratory experiments \cite{aldridge_axisymmetric_1969, lebars_fluid_2021} and have been observed at the surface of the sun and in stars \cite{loptien_globalscale_2018a, gizon_solar_2021, triana_identification_2022, rieutord_spectroscopic_2023}.
Unfortunately, the core of the Earth lies hidden below the mantle, rendering a similar observation of these modes impossible on Earth.
Being mostly kinetic in nature, inertial modes have only a weak magnetic signature. Furthermore, because their frequencies are high (with periods much less than a year), any signal will not only be smoothed by the weakly conducting mantle but will overlap with variations in the external geomagnetic field, making it difficult to isolate.

Another class of modes are torsional (Alfv\'en) modes, first introduced by \cite{braginsky_torsional_1970}. 
They can be understood as perturbations to a Taylor state \cite{taylor_magnetohydrodynamics_1963}, a quasi-steady equilibrium between Coriolis and Lorentz forces.
Torsional modes perturb this equilibrium in the form of differentially rotating geostrophic cylinders, where the magnetic tension through stretching of the background magnetic field that permeates the cylinders acts as the restoring force. Although the geometry of these waves is specific to rotationally-dominant flows, the interplay between the velocity and magnetic field is characteristic of Alfv\'en waves in a non-rotating magnetohydrodynamic fluid \cite{alfven_existence_1942}.
Due to their pseudo-geostrophic structure \cite{gans_hydromagnetic_1971}, a one-dimensional evolution equation can be derived, revealing that the frequency of torsional modes is proportional to the cylindrical average Alfv\'en velocity
\begin{equation}\label{eq:va}
	v_{A}(s) = \sqrt{\frac{1}{4\pi H \mu_0 \rho} \oint\int_{-H}^{H}\,(\BB_0\bcdot\e_s)^2\,\mathrm{d}z\mathrm{d}\phi},
\end{equation}
where $H$ is the half height of the fluid column, $\mu_0$ the magnetic permeability of free space, $\rho$ the fluid density, $\BB_0$ the steady background magnetic field, $\e_s$ the unit vector along the cylindrical radius.

Using identified torsional waves, the relationship between $v_A(s)$ and modal frequency can be used to infer $v_A(s)$, and therefore information on a specific part of the magnetic field itself hidden inside the Earth's core.
Previous estimates of the fundamental torsional wave period were about 60 years \cite{braginsky_torsional_1970,zatman_torsional_1997}, corresponding to $\BB_0\bcdot\e_s \approx 0.5\,\mathrm{mT}$ in the core, but more recently the identification of faster signals has led to an update of the fundamental period to 6 years, corresponding to $\BB_0\bcdot\e_s \approx 3-5\,\mathrm{mT}$ \cite{gillet_fast_2010}, in better agreement with numerical dynamo models.
The analysis of the angular momentum carried by core flows revealed a remarkable correlation in phase and amplitude with the 6-year variation observed in the length-of-day \cite{gillet_fast_2010, gillet_planetary_2015, finlay_gyres_2023}.
Torsional waves only explain a small part of the observed SV, and their direct identification from magnetic measurements or global magnetic field models remains challenging, because their magnetic signature is likely too small to be confidently separated from other dynamics occurring at similar time scales \cite{chulliat_geomagnetic_2014a, cox_observational_2016}.
Instead, the identification of torsional modes within Earth's core relies on the flows obtained by the inversion of geomagnetic field model data. These inversions include additional assumptions about the dynamics of the flow, which can be viewed as a filter of the geomagnetic field data, allowing the isolation of torsional modes within the data.

The final class of modes which we describe are Magneto-Coriolis (MC) modes, which arise when a balance of Coriolis and Lorentz force dominates the momentum budget.
In the literature, often a separation into slow and fast MC modes is presented, where fast MC modes refer to what we have classed as inertial modes (governed by a balance of Coriolis and inertial forces), and slow modes governed by a balance of Coriolis and Lorentz forces. 
Because the fast modes are only weakly influenced by the Lorentz force but strongly influenced by the inertial force, in what follows we exclude them from our classification of MC modes.
Initially discussed by Lehnert \cite{lehnert_magnetohydrodynamic_1954} as plane waves, numerous studies have developed the theory of MC modes.
A variety of studies have considered different geometries from thin spherical shells to full spheres, along with different background fields and forcing terms \cite{braginsky_magnetohydrodynamics_1964, braginsky_magnetic_1967, hide_free_1966 , malkus_hydromagnetic_1967}.
For simplicity, many of these studies not only used a simple  axisymmetric background magnetic field, but also a perfectly conducting boundary condition so that $\BB\bcdot \vec{n} = 0$ at the surface, with $\vec{n}$ the normal vector. 
Neither of these assumptions are representative of the Earth's core, and more recent works have studied MC modes using more appropriate insulating boundary condition for the magnetic field \cite{schmitt_magnetoinertial_2010 , gerick_fast_2021, luo_waves_2022b} and non-axisymmetric background magnetic fields  \cite{vidal_fossil_2019, gerick_fast_2021}. 
Some studies have used non-axisymmetric background magnetic fields with  perfectly conducting boundary conditions, but these calculations cannot be easily related to the Earth \cite{vidal_fossil_2019}.

For a non-axisymmetric poloidal background magnetic field, it was found that some MC modes can have periods corresponding to the interannual period range in Earth's core \cite{gerick_fast_2021}.
In this study, the flow was supposed quasi-geostrophic (QG), appropriate under rapid rotation. 
The quasi-geostrophic Magneto-Coriolis (QGMC) modes at interannual periods combined a small azimuthal wave number with a large cylindrical wave number. 
Near the equator, such a spatial structure projects onto large latitudinal length scales on the core-mantle-boundary (CMB) owing to the local steep gradient of the half height $H$ with $s$.
In this low-latitude region, it was further found that wave-like patterns in core-flows around a period of 7 years showed strong similarities in the phase speed, azimuthal wave number and peak amplitude to the numerically calculated QGMC mode \cite{gillet_satellite_2022}. 

No systematic study has yet investigated the sensitivity of these interannual MC modes on the choice of background magnetic field configuration.
Here, we investigate several poloidal and toroidal background magnetic fields, both axisymmetric and non-axisymmetric. 
We consider all these different choices of background state as they form important contributions to the geomagnetic field, both in the present day but also likely in its general morphology as indicated by high-resolution direct numerical simulations of the geodynamo\cite{aubert_state_2023}.

At high resolution, we compute all modes (i.e. the dense spectra) for each background magnetic field. 
Rather than attempting to study the effect of the background field on all these modes, we investigate the properties of a hierarchy of subsets,  categorized by their observational, geomagnetic and kinematic relevance.

The remainder of this paper is set out as follows. In Section \ref{sec:linearmodel} we introduce the theoretical and numerical background of the linear model used to solve for the eigen modes. 
Section \ref{sec:data} revisits the satellite magnetic field observations over the last two decades and a set of observational and geomagnetic constraints are derived from the data to select only a subset of relevant modes.
The results are presented in Section \ref{sec:results}, before a final discussion in Section \ref{sec:discussion}.

\section{Linear model of the fluid core}\label{sec:linearmodel}

Earths' core is modelled by a spherical core of radius $L$ containing an incompressible, rotating and electrically conducting fluid. The time evolution of the velocity $\UU$ and magnetic field $\BB$ are given respectively by the momentum and induction equation
\begin{subequations}
    \label{eq:mhdeq_dim}
    \begin{align}
        \frac{\partial \UU}{\partial t} + \advection{\UU}{\UU} + 2\rot\times\UU & = -\frac{1}{\rho}\nabla P + \frac{1}{\rho\mu} \left(\curl\BB\right)\times\BB + \nu\bnabla^2\UU + \vec{F}, \\
        \frac{\partial \BB}{\partial t}                                         & = \curl\left(\UU\times\BB\right) + \eta \bnabla^2\BB,
    \end{align}
\end{subequations}
where $\rot$ is the axis of uniform rotation, $\rho$ is the uniform fluid density, $P$ is the reduced hydrodynamic pressure, $\mu$ is the magnetic permeability, $\nu$ is the kinematic viscosity, $\vec{F}$ represents any driving force such as buoyancy, and $\eta$ is the magnetic diffusivity.  

Assuming characteristic scales for length of $L$, magnetic field of $B_0$, time of $t_A = L\sqrt{\rho\mu}/B_0$ (the Alfv\'en time), and velocity of $L/t_A$, equations \eqref{eq:mhdeq_dim} read

\begin{subequations}
    \begin{align}
        \frac{\partial \UU}{\partial t} + \advection{\UU}{\UU} + \frac{2}{\Le}\e_z\times\UU & = -\nabla P + \left(\curl\BB\right)\times\BB + \frac{\Ek}{\Le}\bnabla^2\UU + \vec{F}, \\
        \frac{\partial \BB}{\partial t}                                                     & = \curl\left(\UU\times\BB\right) + \frac{1}{\Lu} \bnabla^2\BB,
    \end{align}
\end{subequations}
where all the vector quantities have been replaced by their non-dimensional versions, and $\Le$, $\Lu$, $\Ek$ are the Lehnert, Lundquist and Ekman numbers, respectively. 
We use both cylindrical coordinates $(s,\phi,z)$, and spherical coordinates $(r,\theta,\phi)$, denoting unit vectors in, for example, the $z$ direction as $\e_z$.

These non-dimensional numbers can be written as the ratios of different time scales \cite{labbe_magnetostrophic_2015}, namely the rotation time $t_\Omega$, the Alfv\'en time $t_A$, the viscous diffusion time $t_\nu$ and the magnetic diffusion time $t_\eta$,

\begin{subequations}
    \begin{align}
        \Le & = \frac{t_\Omega}{t_A} = \frac{B_0}{L\Omega\sqrt{\mu_0\rho}} \sim \mathcal{O}(10^{-4}), \\
        \Lu & = \frac{t_\eta}{t_A} = \frac{B_0L}{\eta\sqrt{\mu_0\rho}} \sim \mathcal{O}(10^{5}),      \\
        \Ek & = \frac{t_\Omega}{t_\nu} = \frac{\nu}{L^2\Omega} \sim \mathcal{O}(10^{-15}).
    \end{align}
\end{subequations}
The orders of magnitudes of these numbers are given as estimates for Earth's core.
We find that the Ekman number is very small at the time scales of interest and so we neglect viscous diffusion in all that follows.

We assume the velocity, magnetic field and pressure evolve as periodic perturbations to a steady background state, so that
\begin{subequations}
    \begin{align}
        \UU(\vec{r},t) & = \UU_0(\vec{r})+ \U(\vec{r}) e^{\lambda t}, \\
        \BB(\vec{r},t) & = \BB_0(\vec{r})+ \B(\vec{r}) e^{\lambda t}, \\
        P(\vec{r},t)   & = P_0(\vec{r})+ p(\vec{r}) e^{\lambda t},
    \end{align}
\end{subequations}
where $\lambda=\mathrm{i}\omega-\sigma$  with $\omega$ the frequency and $\sigma$ the damping rate.
In the following, we consider a steady background magnetic field $\BB_0(\vec{r})$, no forcing ($\vec{F}=\vec{0}$) and zero assumed background flow ($\UU_0(\vec{r}) = \vec{0}$). 

The linearised set of equations governing the perturbation are:
\begin{subequations}
    \label{eq:mhdeq_lin_nondim}
    \begin{align}
        \lambda\U + \frac{2}{\Le}\e_z\times\U & = -\nabla p + \left(\curl\B\right)\times\BB_0+ \left(\curl\BB_0\right)\times\B,          \\
        \lambda\B                             & = \curl\left(\U\times\BB_0\right) + \frac{1}{\Lu} \bnabla^2\B.
    \end{align}
\end{subequations}

The  fluid volume of radius $r=1$ is denoted $\mathcal{V}$, its boundary as $\partial \mathcal{V} $ and its exterior $\hat{\mathcal{V}}$ (so that $\mathbb{R}^3=\mathcal{V}\cup\hat{\mathcal{V}}$). 
We can then project the evolution equations as
\begin{subequations}
    \label{eq:energy_equations}
\begin{align}
    \lambda\left<\tilde{\U},\U\right>_\mathcal{V} &= \left<\tilde{\U}, \frac{2}{\Le}\e_z\times\U + \curl\B\times\BB_0 + \curl\BB_0\times\B \right>_\mathcal{V},\label{eq:ekin_c}\\
    \lambda\left<\tilde{\B},\B\right>_{\mathbb{R}^3} &= \left<\tilde{\B}, \curl\left(\U\times\BB_0\right) + \frac{1}{\Lu}\bnabla^2\B\right>_{\mathbb{R}^3},\label{eq:emag_c}
\end{align}
\end{subequations}
where
\begin{align}
    \left<\vec{v},\vec{w}\right>_\mathcal{V} =  \int_\mathcal{V}\vec{v}^*\bcdot\vec{w}\,\mathrm{d}V.
\end{align}
In the above, $\vec{v}^*$ denotes the complex conjugate of $\vec{v}$, and $\tilde{\U}$ and $\tilde{\B}$ are velocity and magnetic test functions respectively.
The kinetic energy equation \eqref{eq:ekin_c} requires only the interior volume integral over $\mathcal{V}$, since $\U = \vec{0}$ in $\hat{\mathcal{V}}$. 
We note that the projection of the pressure gradient is omitted in \eqref{eq:ekin_c}, as it vanishes for an incompressible velocity field. The projection \eqref{eq:emag_c} is defined over all space for numerical expediency.

\subsection{Galerkin method and bases}

We discretise equations \eqref{eq:energy_equations} by expressing both the flow and magnetic field as linear combinations of basis vectors,
\begin{subequations}
\begin{align}
    \U = \sum_{i=1}^{\operatorname{dim}(\boldsymbol{\mathcal{P}}^\U_N(\mathcal{V}))} \alpha_i\U_i,\\
    \B = \sum_{i=1}^{\operatorname{dim}(\boldsymbol{\mathcal{P}}^\B_N(\mathbb{R}^3))} \beta_i\B_i,
\end{align}
\end{subequations}
with complex coefficients $\alpha_i,\beta_i \in \mathbb{C}$ and  the respective subspaces of the flow and magnetic field, $\boldsymbol{\mathcal{P}}^\U_N(\mathcal{V})$ and $\boldsymbol{\mathcal{P}}^\B_N(\mathbb{R}^3)$, which are introduced and defined in detail in the following. Discretised versions of the dynamical equations are formed in \eqref{eq:energy_equations} by back-projecting onto the same subspaces. 
The subspaces are chosen to have the following expedient properties:
\begin{itemize}
    \item Each basis vector is a geometrically admissible solution, in that it satisfies all conditions related to boundaries and differentiability.
    \item Although defined in terms of spherical polar coordinates, the bases have a Cartesian homogeneous complexity $N$.
    \item Each basis vector is built from a spherical harmonic and a terse combination of Jacobi polynomials, both of which are spectrally convergent.
    \item Certain projections are optimally sparse, which reduces memory requirements.
\end{itemize}

Because the geometry is spherical, it is easiest to define the subspaces in this geometry. However, it is helpful to express the spaces in terms of Cartesian coordinates in order that we can define a homogeneous measure of the spatial complexity $N$ (that is, invariant under rotation).
The subspaces are defined as follows: 

\begin{subequations}
\begin{align}
    \boldsymbol{\mathcal{P}}^\U_N(\mathcal{V})       & = \lbrace \U \in \mathcal{P}_N^3\, \big|\, \bnabla\bcdot\U = 0\, \text{ in }\, \mathcal{V}, \U\bcdot\vec{n} = 0\,\text{ on }\, \partial \mathcal{V} \rbrace, \\
    \boldsymbol{\mathcal{P}}^\B_N\left(\mathbb{R}^3\right)       & = \begin{cases}
        \B \in \mathcal{P}_N^3 \big|\, \bnabla{\bcdot}\B = 0 & \text{ in } \mathcal{V},\\
        \hat{\B} \in \mathcal{I}_N^3 \big|\, \bnabla{\bcdot}\hat{\B} = 0, \hat{\B} = -\nabla \Phi, \Phi \in \mathcal{I}_N & \text{ in } \hat{\mathcal{V}},\\
        \B = \hat{\B} & \text{ on } \partial\mathcal{V},
    \end{cases}
\end{align}
\end{subequations}
where $\mathcal{P}_N = \lbrace x^iy^jz^k\, \big| i,j,k \in \mathbb{Z}^{0+}, \, 0{\leq}i{+}j{+}k{\leq}N\rbrace $, $\Phi$ is the magnetic potential field in the exterior domain, and $\mathcal{I}_N = \lbrace x^iy^jz^k\, \big| i,j,k \in \mathbb{Z}^{-}, \, 0{\leq}|i{+}j{+}k|{\leq}N\rbrace $.
The subspaces for the flow and magnetic field in $\mathcal{V}$ are built from vectors whose Cartesian coordinates belong to the set $\mathcal{P}_N$, homogeneous multinomials of degree at most $N$. They further satisfy zero divergence, and either impenetrable or electrically insulating boundary conditions\cite{lebovitz_stability_1989, ivers_enumeration_2015, luo_inviscid_2021}. 
We have $\operatorname{dim}(\boldsymbol{\mathcal{P}}^\U_N(\mathcal{V})) = N(N+1)(2N+7)/6$ \cite{ivers_enumeration_2015} and $\operatorname{dim}(\boldsymbol{\mathcal{P}}^\B_N(\mathcal{V})) = N(N-1)(2N+5)/6$. 
The subspace for magnetic field in  $\hat{\mathcal{V}}$ is also homogeneous but built from multinomials with negative exponents.
Because the spherical boundary conditions couple the Cartesian components together, these spaces are compiled from expedient representations in spherical polar coordinates, involving spherical harmonics and polynomials in $r$.
Using such a truncation, sometimes referred to as a triangular truncation as the maximum radial polynomial index decreases with spherical harmonic degree $l$ \cite{holdenriedchernoff_surface_2020, luo_inviscid_2021}, is advantageous.
Not only is the solution space homogeneous in resolution, but certain special classes of solution like the inviscid inertial modes are complete within this space \cite{ivers_enumeration_2015}.
This example is discussed in the Supplementary Material Section S1, comparing a uniform truncation (where the radial degree $n$ is truncated at the same degree for all spherical harmonic degrees $l$) to the triangular truncation.

\subsubsection{Velocity basis}

Since we require $\bnabla\bcdot\U = 0$, we use a classical poloidal-toroidal decomposition of the velocity:
\begin{align}
    \U = \sum_i \alpha_i\U_i = \sum_{l,m,n} \alpha^P_{lmn}\vec{P}_{lmn} + \sum_{l,m,n} \alpha^Q_{lmn}\vec{Q}_{lmn},
\end{align}
with $\alpha^P_{lmn}, \alpha^Q_{lmn} \in \mathbb{C}$.
The vectors, written in spherical coordinates $(r,\theta,\phi)$, are
\begin{subequations}
    \label{eq:poltor_pq}
    \begin{align}
        \vec{P}_{lmn} & = \curl\curl P_{ln}(r)Y_l^m(\theta,\phi)\vec{r}, \\
        \vec{Q}_{lmn} & = \curl Q_{ln}(r) Y_l^m(\theta,\phi)\vec{r}. 
    \end{align}
\end{subequations}
Here, $Y_l^m(\theta,\phi)$ is the (fully normalized) spherical harmonic of degree $l$ and order $m$, and $\vec{r}=r\e_r$. 
The toroidal and poloidal scalar functions $Q_{ln}$ and $P_{ln}$, where $n$ is the radial index, are chosen to satisfy the appropriate boundary condition at the surface (here $r=1$), orthogonality, and regularity at the origin \cite{livermore_quasilpnorm_2010, livermore_galerkin_2010}.
The assumed inviscid fluid only satisfies a condition of impenetrability on $r=1$, leaving the toroidal scalar function $Q_{ln}$ unconstrained but requiring the poloidal scalar function to satisfy
\begin{align}
    P_{ln}(1) = 0.
\end{align}
In addition, to satisfy the regularity of the velocity at the origin ($r=0$), these functions need to take the form $Q_{ln} \sim P_{ln} \sim r^l f(r^2)$ where $f$ is an arbitrary polynomial.

A set of poloidal and toroidal scalar functions satisfying these conditions is presented in Appendix \ref{app:velocitybasis}, where we give the analytical expressions of the inner product and the projections onto the Coriolis term. They are all built from terse expansions in Jacobi polynomials \cite{livermore_galerkin_2010}.

\subsubsection{Magnetic field basis}

In the same way as the velocity, we write the magnetic field in the interior volume $\mathcal{V}$ as a linear combination of poloidal and toroidal vectors
\begin{align}\label{eq:mag_basis}
    \B & = \sum_i \beta_i\B_i =  \sum_{l,m,n} \beta^S_{lmn}\vec{S}_{lmn} + \sum_{l,m,n} \beta^T_{lmn}\vec{T}_{lmn},
\end{align}
with $\beta^S_{lmn}, \beta^T_{lmn}  \in \mathbb{C}$.
Analogous to \eqref{eq:poltor_pq}, the vectors are written as
\begin{subequations}
    \label{eq:poltor_st}
    \begin{align}
        \vec{S}_{lmn} & = \curl\curl S_{ln}(r)Y_l^m(\theta,\phi)\vec{r}, \\
        \vec{T}_{lmn} & = \curl T_{ln}(r) Y_l^m(\theta,\phi)\vec{r}.
    \end{align}
\end{subequations}
Assuming the overlying mantle to be insulating, the magnetic field is required to match its three components to a potential field $-\nabla \Phi$ on $r=1$. 
The expression of the poloidal field $\vec{S}_{lmn}$ in the interior matches to a field $\hat{\vec{S}}_{lmn}$ in the exterior $\hat{\mathcal{V}}$, written as
\begin{align}
    \hat{\vec{S}}_{lmn} = -l S_{ln}(1) \nabla I_m^l,
\end{align}
with $I_m^l = r^{-(l+1)}Y_m^l$, so that the associated magnetic potential field $\Phi = l S_{ln}(1) I_m^l$.
The continuity across the surface is equivalent to the conditions
\begin{subequations}\label{eq:insulatingBC}
\begin{align}
    T_{ln}(1)                                         & = 0,                \\
    \frac{\partial S_{ln}(r)}{\partial r}\bigg|_{r=1} & = -(l+1) S_{ln}(1). \label{eq:insulatingBCpol}
\end{align}
\end{subequations}
Again, to satisfy the regularity of the magnetic field at the origin, $T_{ln} \sim S_{ln} \sim r^l f(r^2)$.

A set of functions satisfying these conditions is given in Appendix \ref{app:insulatingmfbasis}, together with the analytical expressions of the inner product, as well as the projection onto the vector Laplacian.
The basis we have chosen is orthogonal w.r.t the projection onto the vector Laplacian, and tridiagonal w.r.t the inner product.
In doing so, we also find that all other projections in the induction and momentum equation are banded in the radial degree as well (for the orthogonal inviscid velocity basis considered here).
The bandwidth of the induction term $\curl(\U\times\BB_0)$ and Lorentz term depends on the radial degree $n$ of the background magnetic field (or flow, if considered).
This property is desirable for increased resolution and problem sizes, needed to resolve modes of complex background magnetic field structures.

\subsubsection{Magnetic induction and Lorentz force projections}

An important part of equations \eqref{eq:energy_equations} we need to solve are the projections 
\begin{align}
	\left<\tilde{\U}, \left(\curl\B\right)\times\BB_0 + \left(\curl\BB_0\right)\times\B \right>_\mathcal{V},\\
    \left<\tilde{\B}, \curl\left(\U\times\BB_0\right)\right>_{\mathbb{R}^3},
\end{align} 
for each choice of flow and magnetic field basis vector. Based on the seminal work of \cite{bullard_homogeneous_1954a} and \cite{ivers_scalar_2008}, these projections involve both radial integrals and integrals over a spherical surface, the latter of which can be written as Adam-Gaunt and Elsasser integrals \cite{bullard_homogeneous_1954a, james_adams_1973}
\begin{align}
    A_{ijk} & = \oint\int Y_iY_jY_k\sin\theta\,\mathrm{d}\theta\mathrm{d}\phi,                                                                                                                                             \\
    E_{ijk} & = \oint\int Y_k \left(\frac{\partial Y_i}{\partial \theta}\frac{\partial Y_j}{\partial\phi} - \frac{\partial Y_i}{\partial\phi}\frac{\partial Y_j}{\partial \theta}\right) \,\mathrm{d}\theta\mathrm{d}\phi,
\end{align}
where we have abbreviated the notation of the spherical harmonics so that $Y_i = Y_{l_i}^{m_i}$.
What remains to be calculated of the projections are 1D equations, only depending on the radius $r$, for each combination of the basis vectors.
The detailed equations are shown in Appendix \ref{app:li_projections}.
As the induction equation is integrated over $\mathbb{R}^3$, some additional terms arise in the projections if $\U\neq\vec{0}$ on the surface, even though $\U=\vec{0}$ in the exterior $\hat{\mathcal{V}}$. The details are discussed in Appendix \ref{app:external}.
The Adam-Gaunt and Elsasser variables $A_{ijk}$ and $E_{ijk}$ are calculated numerically through Wigner-symbols using the \texttt{WIGXJPF} library \cite{johansson_fast_2016}.
The integrals over $r$ are exactly calculated using Gauss-Legendre quadrature.

\subsection{Eigen problem}\label{subsec:eigen}

The discretised projected version of the momentum and induction equations \eqref{eq:energy_equations}, i.e. the projection onto the basis elements $\U_i$ and $\B_i$, then can be written as the generalised eigenproblem:
\begin{equation}\label{eq:gep}
    \lambda \vec{M}\vec{x} = \vec{N}\vec{x},
\end{equation}
with
\begin{align}
    \vec{M} = \begin{pmatrix}
                  \vec{V} & \vec{0} \\
                  \vec{0} & \vec{W}
              \end{pmatrix}, \\
    \vec{N} = \begin{pmatrix}
                  \vec{C} & \vec{L} \\
                  \vec{A} & \vec{D}
              \end{pmatrix}.
\end{align}
The submatrices $\vec{V}$ and $\vec{W}$ arise from inner products of the velocity basis with itself, and the magnetic field basis with itself, respectively. The submatrices $\vec{C}$ and $\vec{L}$ are respectively the projections of the Coriolis and Lorentz terms onto the velocity basis, while $\vec{A}$ and $\vec{D}$ are the projections of the induction and the magnetic diffusion terms onto the magnetic field basis.
The matrices $\vec{M}$ and $\vec{N}$, each of size $\mathcal{S} \times \mathcal{S}$, are sparse and for the bases considered here, $\vec{M}$ is symmetric tridiagonal.
If we consider the full subspaces $\boldsymbol{\mathcal{P}}^\U_N(\mathcal{V}))$ and $\boldsymbol{\mathcal{P}}^\B_N(\mathcal{V}))$, i.e. without any symmetry assumptions to reduce the problem size, we have $\mathcal{S} = N(1+6N+2N^2)/3$.
An eigensolution $(\lambda,\vec{x})$ is the solution to \eqref{eq:gep}, with $\lambda$ the complex eigenvalue and the complex eigenvector $\vec{x}$ containing a list of coefficients $[\alpha^P_{lmn}, \alpha^Q_{lmn}, \beta^S_{lmn}, \beta^T_{lmn}]$.
For small polynomial degrees $N \lesssim 40$, the matrix size $\mathcal{S}$ is sufficiently small that we can solve \eqref{eq:gep} directly, using a dense generalised Schur factorisation, giving access to the dense spectrum of modes.
Keeping the matrix size $\mathcal{S}$ moderate but increasing $N$ requires exploitation of symmetry, in order to reduce the number of angular modes involved.
In the most general case, if $\BB_0$ has no particular symmetry, all angular modes are coupled together. 
However, if $\BB_0$ has particular properties such as equatorial symmetry or axial axisymmetry, then the solution  separates into independent symmetry classes which can be investigated in separate calculations \cite{gubbins_symmetry_1993}. 
In particular, under axisymmetry, each azimuthal wave number $m$ can be considered individually. 
For such $\BB_0$, the upshot is that number of angular modes, for any calculation, can be vastly reduced.

For large matrix sizes $\mathcal{S}$, computing the dense spectrum becomes infeasible due to computational effort and memory requirements, and instead we can turn to iterative methods that exploit matrix sparsity to calculate a subspace of eigensolutions.
One of such method is the implicitly restarted Arnoldi method, available in the \texttt{ARPACK} library \cite{lehoucq_arpack_1998}.
Iterative methods are good at finding a few eigenvalues with extremal properties, for example, largest or smallest in amplitude or in real/imaginary part.
To find eigensolutions that lie within the spectrum and not at the extremes, we can use a shift-invert method, where we shift the spectrum around the target $\lambda_t$, so that
\begin{equation}\label{eq:shiftinvert}
    \frac{1}{\lambda-\lambda_t}\vec{x} = \left(\vec{N}-\lambda_t\vec{M}\right)^{-1}\vec{M}\vec{x} = \vec{M}_t\vec{x}
\end{equation}
which defines a new linear operator $\vec{M}_t$.
Written is this way, the eigenvalues $1/(\lambda - \lambda_t)$ are largest, when $\lambda$ is close to $\lambda_t$.
The right-hand side can be calculated without explicitly calculating the inverse $(\vec{N}-\lambda_t\vec{M})^{-1}$ through the use of the sparse LU factorisation provided for example by \texttt{UMFPACK} \cite{davis_algorithm_2004} and \texttt{Intel MKL Pardiso} \cite{schenk_efficient_2000}.
Using the shift-invert method allows tracking of a given eigensolution through parameter space and different numerical resolutions.

Any solution of either a dense or shift-invert calculation needs to be checked for convergence in resolution. 
We deem a numerical eigensolution to be \emph{numerically relevant} if it is converged as judged by the following criteria:
\begin{itemize}
    \item The frequency of the mode should not change its value more than 10\% between an eigensolution calculated at truncation $N$ and one at $N+2$. For some modes that are further analysed, we verify stricter convergence by tracking individual modes up to higher resolutions and verifying the eigenvalues converge towards a finite value.
    \item Between two resolutions, $N$ and $N+2$, the eigenvectors must correlate to a minimum threshold ($0.99$). The correlation is performed by appropriate padding with zeros of the eigenvector at the lower resolution.
    \item Another requirement is spectral convergence. We consider the spectral energy density as calculated as a function of Cartesian monomial degree $\tilde{n}$ which takes account of structure in all spatial directions; this is distinct from considering the spectrum only as a function of $l$ which ignores radial complexity. We calculate this by exploiting the fact that every basis vector has a particular Cartesian complexity, binning the squared magnitude of the coefficients at each $\tilde{n}$. We calculate the energy at the peak, and compare this to the energy at the highest resolution $N$.
    
    For solutions with an axisymmetric $\BB_0$, when each $m$ can be considered independently, we can use a higher resolution and we can be stricter: for these cases we require the energy density at the truncation to be 1\% of the peak energy density. For the solutions with a non-axisymmetric $\BB_0$, we weaken this to a factor of 5\%, due to the restrictions in resolution for dense calculations. We then confirm further spectral decay by tracking relevant solutions to a higher resolution.
    
\end{itemize}

\section{Interannual secular variation and hierarchical classification}\label{sec:data}

In a typical calculation, the matrix size $\mathcal{S}$ might be several thousand, leaving typically many hundreds of modes \emph{numerically relevant}. Many of these modes will be invisible to observation, either because they are too rapid (so their time-dependence is smoothed out by the weakly conducting mantle), or their lengthscale is too short (and so their signature is lost in small scale unmodeled signals). Even if modes have an appropriate frequency and lengthscale, they may not describe the present-day interannual pattern of SV in Earth (although they may have been present in Earth before modern era of observation).
We therefore introduce a system of mode classifaction, with specific criteria described below. Broadly, the classes are:
\begin{enumerate}
    \item A mode is \emph{numerically relevant} if it is a converged solution of the equations \eqref{eq:energy_equations}.
    \item A mode is \emph{observationally relevant} if in principle it would be observable in the currently available 24 years of continuous satellite data.
    \item A mode is \emph{geomagnetically relevant} if it is consistent with the structure of interannual SV over the last 24 years.
    \item A mode is \emph{kinematically
    relevant} if it bears resemblance to other core-flow inversions based on recent observations of the magnetic field.
\end{enumerate}

This classification is also hierarchical. For example, a mode which is \emph{geomagnetically relevant} is also \emph{observationally relevant} and \emph{numerically relevant}. 

The classification presented here is configured to the geomagnetic observations currently available from satellites. 
We want to highlight that the details of the classification laid out below are not necessarily universal but can easily be adapted to apply to other datasets, e.g. archaeomagnetic observations or the data from future satellite missions.

\subsection{Interannual secular variation}

\begin{figure}
    \centering
    \begin{minipage}{0.49\textwidth}
       (a)

        \includegraphics[width=\linewidth]{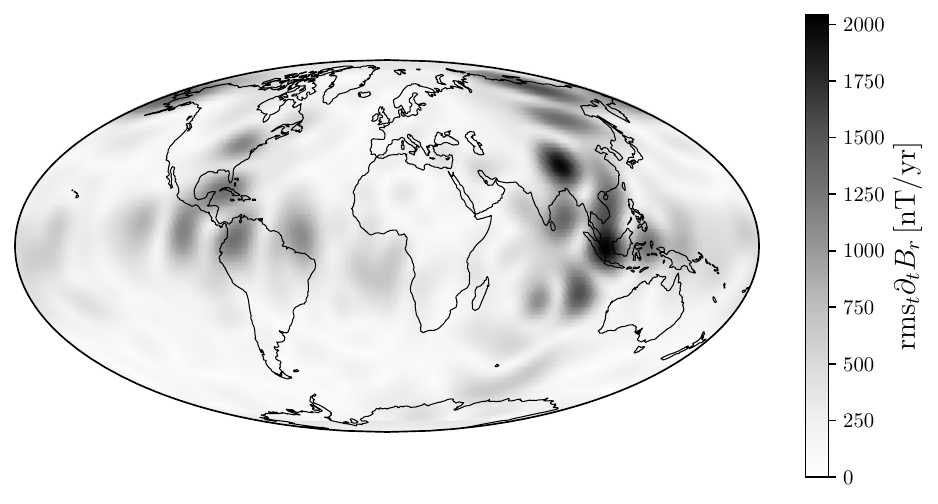}
        
       (c)
       
        \includegraphics[width=\linewidth]{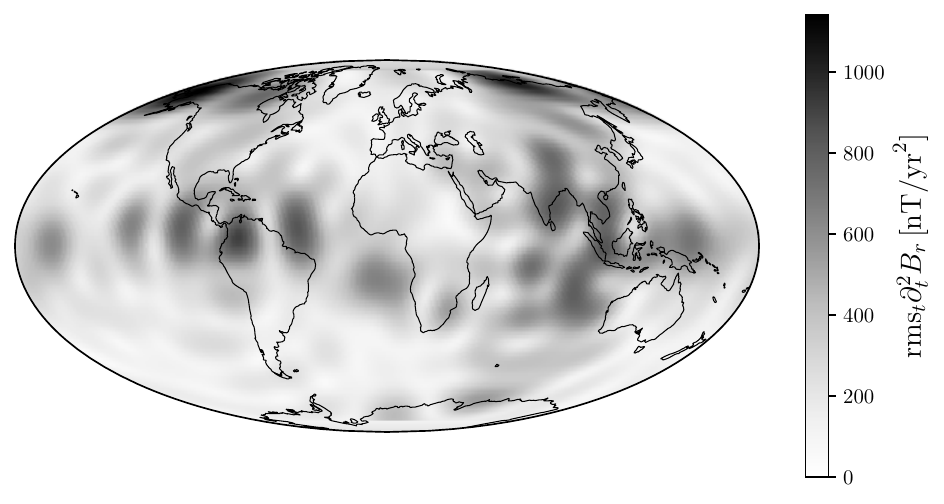}
    \end{minipage}
    \begin{minipage}{0.49\textwidth}
    (b)
    
    \includegraphics[width=\linewidth]{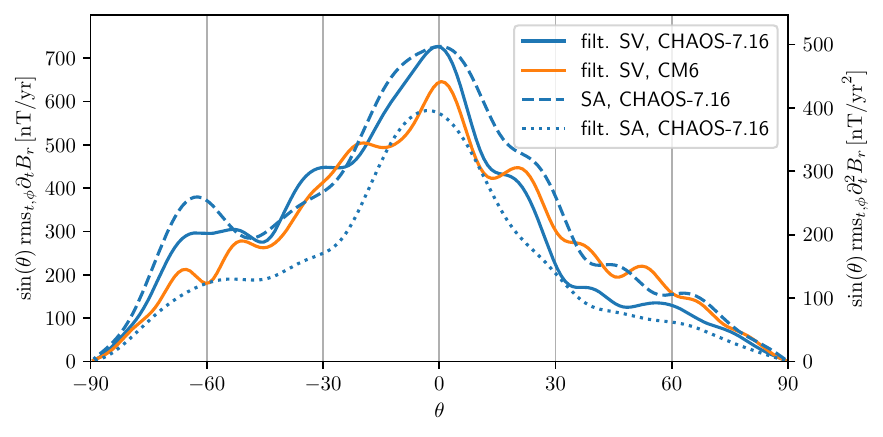}

    (d)
    
    \includegraphics[width=\linewidth]{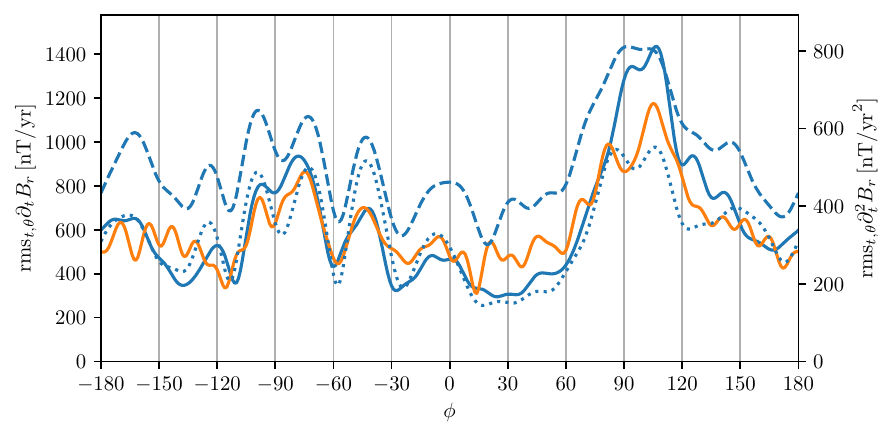}
    \end{minipage}
    
    \caption{Properties of interannual geomagnetic change 1999-2023. (a) Temporal rms of radial SV, band-pass filtered around 1--23.7 yr, at the CMB from CHAOS-7.16. (b) Temporal and longitudinal rms as a function of latitude $\theta$ for filtered SV and SA based on CHAOS-7.16 and CM6. Each profile is weighted with the area factor of $\sin\theta$. (c) Temporal rms of radial SA at the CMB from CHAOS-7.16. (d) Temporal and latitudinal rms as a function of longitude $\phi$ for filtered SV and SA, with the same legend as (b). Contours of the continents are projected down to the CMB for reference.}
    \label{fig:geomagnetic_observations}
\end{figure}

Within state-of-the-art global geomagnetic field models, the SV is available up to spherical harmonic degree $l=17$, whereas the secular acceleration (SA) is reliable up to degree $l=10$, considered here \cite{lesur_rapid_2022, finlay_gyres_2023}.
In Figure \ref{fig:geomagnetic_observations} we illustrate the SV and SA extracted from the CHAOS-7.16 model by their temporal root mean square (rms), \cite{finlay_chaos7_2020}, as well as the CM6 model \cite{sabaka_cm6_2020}. 
We use the \texttt{chaosmagpy} package to process the model coefficients \cite{kloss_ancklo_2023}.
The times 01/2000--08/2023 are considered for the CHAOS-7.16 model and the times 01/2000--12/2018 for the CM6 model.
In order to remove variations occurring at time scales longer than the observational time series available, we apply a band-pass filter of 1--23.7~yr.
The upper bound of 23.7~yr is determined by the length of the CHAOS-7.16 dataset. 
The lower bound of one year is chosen to avoid contamination through external signals and it is believed to be about the shortest observable period for signals of internal origin due to the slightly conducting mantle \cite{backus_foundations_1996}.
Another way of describing the dynamics at interannual timescales is by the SA, giving more emphasis to the most rapid dynamics.
We compare the unfiltered and filtered SA to the filtered SV and find qualitatively similar features.

The main characteristic of secular changes, on interannual timescales, is a focus of radial SV and SA in the equatorial region. 
This is clearly visible in Figure \ref{fig:geomagnetic_observations}(a, c), highlighting the equatorial structure and is further quantified in the rms taken both over time and longitude in Figure \ref{fig:geomagnetic_observations}(b).
In both the filtered SV and SA data a clear peak in the rms is found at the equator, $\theta=0$. 
Some north-south asymmetry is observed, with a stronger rms near $\theta=-60^\circ$ compared to the southern hemisphere. 
This local peak in the rms field is removed, when filtering the data at shorter periods (see Figure S1 in the suppl. material).
Lastly, we show the rms over time and latitude as a function of longitude $\phi$ in Figure \ref{fig:geomagnetic_observations}(d). 
The rms field as a function of longitude is more complex in its structure, showing several peaks in amplitude, with stronger amplitudes around Asia ($\phi=90^\circ$) and the Americas ($\phi=-90$).

\subsection{Observationally relevant modes}

Given these basic properties of the interannual SV over the last two decades, we define a mode to be \emph{observationally relevant} with respect to the presently available satellite magnetic observations if

\begin{itemize}
        \item The dominant length-scale of the mode at the surface should be large enough to be observable in the geomagnetic data. Here, this length-scale is defined by the spherical harmonic degree, which should be $l\leq 17$ (in SV data) or $l\leq 10$ (in SA data) based on the current geomagnetic models considered. For any mode, because of the assumed time-dependence, the spatial structure of the mode for any time-derivative are the same. We therefore can use the spatially highest resolution constraint, and so require that the peak amplitude in the poloidal magnetic field component of the mode (which determines $b_r$ at the surface) has to be at a degree $l,|m|\leq 17$. 
        \item The radial length-scale within the core is not constrained by observations. However, in order to focus our attention to the largest scale modes, we assume the peak amplitude in the poloidal magnetic field component should be at a radial degree $n\leq l/2 = 17/2$. This choice is motivated by the uniform truncation in the Cartesian monomial degree $\tilde{n}\sim 17$, i.e. we require the smallest observable length-scale to be equal in all spatial directions.
        \item The quality factor of the mode, $Q = |\omega/2\sigma|$, should be larger than unity in order for the mode to propagate before being damped and therefore to be observable. 
        \item The period of the mode $t=2\pi/|\omega|$ should be interannual, or within observational and geophysical limits. The shortest period is determined by the filter through the slightly conducting mantle, as well as the masking of the internal signal by the ionospheric signal. The longest period is determined by the satellite data availability. We restrict the range of observable periods of the modes to be ${0.5t_A<t<11t_A}$, corresponding to periods between 1.1--24 yr for $t_A = 2.2$~yr. 
        In terms of the frequencies of the numerically calculated modes, given as Alfv\'en frequencies, this requires $0.57 < |\omega| < 12.6$.
    \end{itemize}
    
    \subsection{Geomagnetically relevant modes} 
    The spatial structure of the mode's SV should be in agreement with the rms SV derived from the geomagnetic observations. Qualitatively, this means that most of the mode's magnetic field variations should occur at low latitudes, close to the equator.
    
    \begin{itemize}
        \item  For a \emph{geomagnetically relevant} mode, we require that the temporal and  longitudinal rms of the mode should correlate (using a lower threshold of $0.8$) with the similar profile using the SV filtered between 1--23.7~yr (Figure \ref{fig:geomagnetic_observations}(b), blue line), derived from the CHAOS-7.16 model.
        Each of these profiles is weighted by the factor $\sin\theta$ to take account of the variation of the element of area, $\sin\theta \mathrm{d}\theta \mathrm{d}\phi$ on the spherical surface.
    \end{itemize}
    
  \subsection{Kinematically relevant modes}
Other studies, based on similar geomagnetic datasets to those we have described, have reconstructed core-flows, which generally show a focusing of the azimuthal velocity $u_\phi$ near the equator \cite{kloss_timedependent_2019, istas_transient_2023, ropp_midlatitude_2023}, comparable to the focusing observed in the SV and SA. 
    \begin{itemize}
        \item For a mode to be \emph{kinematically relevant}, we require that the peak amplitude in $u_\phi$ lies at $|\theta|< 30^\circ$.
    \end{itemize}

We want to highlight that the definition of kinematic relevance is only a very crude way of comparing the calculated modes to core-flows obtained from inversions of geomagnetic data.
This comparison is only to put our modes into context of previous works, and is not intended as a geophysical constraint in general.
We do not want to put any constraint on the flow component of the solution, but rather only on the magnetic component which is directly constrained by geomagnetic data.

\section{Results}\label{sec:results}

\subsection{Background magnetic fields and convergence}

We fix the Lehnert number to be $\Le=2\cdot 10^{-4}$. 
For a density $\rho = 1.1\cdot 10^4$~kg/m$^3$, a length-scale $L=3480$~km, this corresponds to $B \approx 6$~mT, $t_A \approx 2.2$~yr.
We use a Lundquist number of $\Lu = 2\cdot 10^4$, corresponding to $\eta \approx 8.8$~m$^2$/s.
This value of $\eta$ is slightly smaller than the value expected for Earth's core, 
although higher magnetic diffusion aids numerical convergence.

\begin{figure}
    \centering
    \begin{minipage}{0.49\textwidth}
        \includegraphics[width=\linewidth]{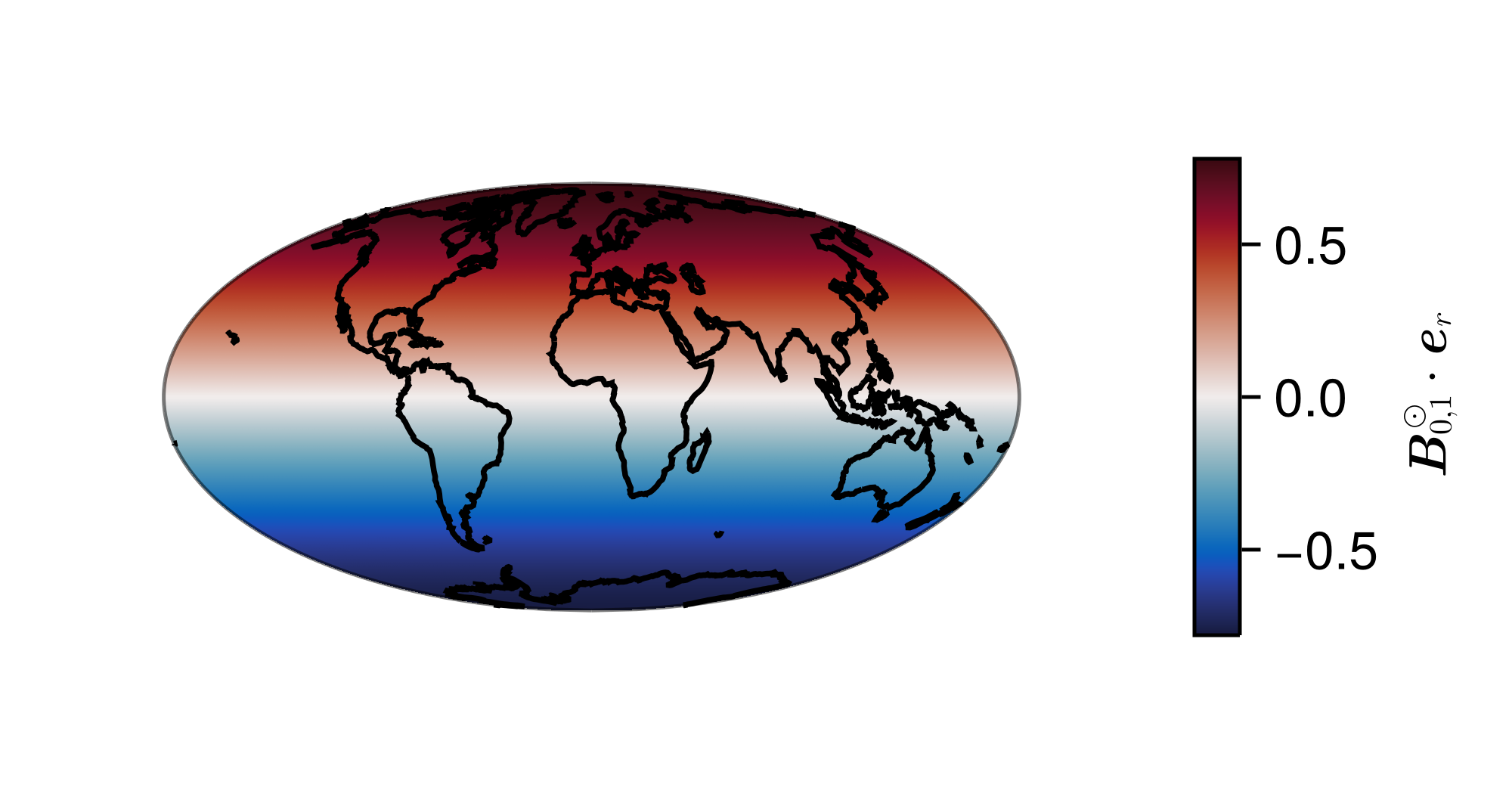}
    \end{minipage}
    \hfill
    \begin{minipage}{0.49\textwidth}
        \includegraphics[width=\linewidth]{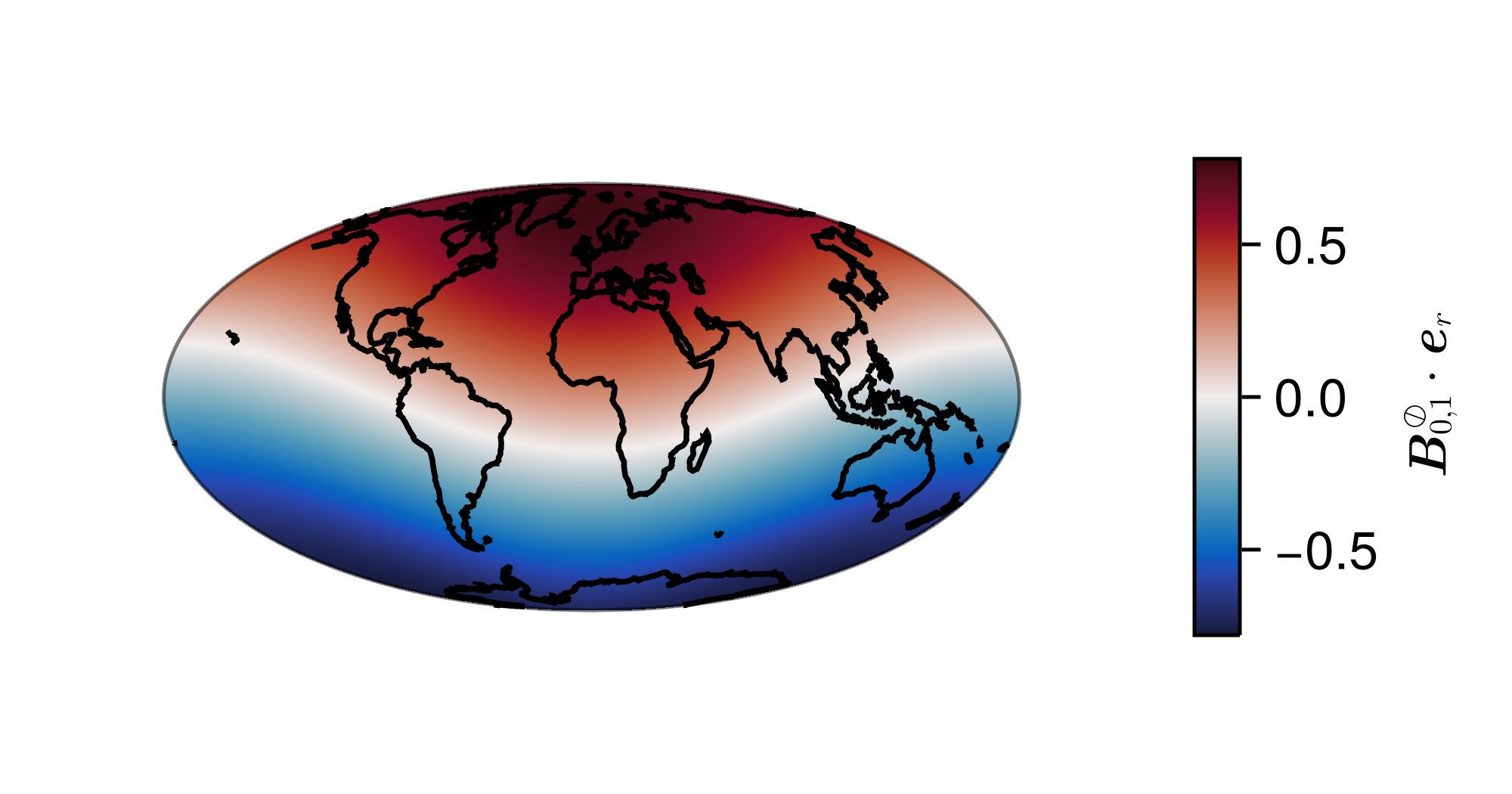}
    \end{minipage}
    \caption{Radial component of the axisymmetric background magnetic field $\BB_{0,1}^\odot$ (left) and $\BB_{0,1}^\oslash$ (right) at the surface of the core. Contours of the continents are projected down to the CMB for reference.}
    \label{fig:B0s_surface}
\end{figure}

Several background magnetic fields, both axisymmetric and non-axisymmetric, constructed from either single or several poloidal and toroidal components are investigated.
The exact expressions and naming conventions are introduced in Table \ref{table:b0new}.
The radial component at the surface of the core of the axisymmetric field $\BB_{0,1}^\odot$ and the non-axisymmetric field $\BB_{0,1}^\oslash$ are illustrated in Figure \ref{fig:B0s_surface}.
By allowing non-axisymmetry in the background state we are able to approximate the Earth's steady magnetic field more accurately at the cost of increased computational effort.
We normalise each background magnetic field to have a unit rms value within the core volume,
\begin{equation}
    B_0 = \frac{1}{V}\int_\mathcal{V} \BB_0\bcdot\BB_0\,\mathrm{d}V = 1,
\end{equation}
with $V = 4\pi/3$. In order to construct a background field that is real, a single magnetic field component is given as the following sum of complex-valued constituents:
\begin{align}
	\BB_{lmn}^{\vec{S}} = \begin{cases}
			\frac{1}{\sqrt{2}}\left((-1)^m\vec{S}_{lmn}+\vec{S}_{l(-m)n}\right), & m>0,\\
            \vec{S}_{lmn}, & m=0, \\
			\frac{\mathrm{i}}{\sqrt{2}}\left(\vec{S}_{lmn}-(-1)^m\vec{S}_{l(-m)n}\right), & m<0,
	\end{cases}
\end{align}
and analogously for a toroidal magnetic field component $\BB_{lmn}^{\vec{T}}$.

\begin{table}
	\caption{Background magnetic field configurations considered in this study. Axisymmetry and non-axisymmetry are annotated by $\odot$ and $\oslash$, respectively.}
	\label{table:b0new}
	\begin{tabular}{lll}
	\hline
	Name & Components & Expression\\
	\hline\\[-1.0em]
    $\BB_{0,1}^\odot$ & $\BB_{101}^\vec{S}$ & ${\curl}{\curl} \frac{1}{2} \sqrt{\frac{7}{46}} f^s_{101}\vec{r}$\\ 
    $\BB_{0,2}^\odot$ & $\BB_{101}^\vec{S},\BB_{101}^\vec{T}$ & $\frac{1}{\sqrt{2}}\BB_{0,1}^\odot + {\curl} \frac{3}{4} \sqrt{\frac{35}{2}}  f^t_{101} \vec{r}$\\
    $\BB_{0,3}^\odot$ & $\BB_{101}^\vec{S},\BB_{201}^\vec{S}$ & $\frac{1}{\sqrt{2}}\BB_{0,1}^\odot + {\curl}{\curl} \frac{1}{32} \sqrt{\frac{5}{13}} f^s_{201}\vec{r} $\\
     $\BB_{0,4}^\odot$ & $\BB_{101}^\vec{S},\BB_{104}^\vec{S}$ & $\frac{1}{\sqrt{2}}\BB_{0,1}^\odot{+}{\curl}{\curl}\frac{1}{256} \sqrt{\frac{1615}{6}} r \left(r^2{-}1\right)^2 \left(195 r^4{-}182 r^2{+}35\right) \cos (\theta )\vec{r} $\\
     $\BB_{0,5}^{\odot,\dagger}$ & $\BB_{101}^\vec{T}$ & ${\curl} \frac{3}{4} \sqrt{35} f^t_{101} \vec{r}$\\
     $\BB_{0,6}^{\odot,\dagger}$ & $\BB_{201}^\vec{S}$ & ${\curl}{\curl} \frac{1}{16} \sqrt{\frac{5}{26}} f^s_{201}\vec{r} $\\[1em]
     \hline \\
     $\BB_{0,1}^\oslash$ & $\BB_{101}^\vec{S},\BB_{111}^\vec{S}/3$ & ${\curl}{\curl} \frac{1}{4} \sqrt{\frac{7}{115}}\left(3f^s_{101} +  f^s_{111}\right)\vec{r}$\\
     $\BB_{0,2}^\oslash$ & $\BB_{101}^\vec{S},\BB_{111}^\vec{S}/3,\BB_{111}^\vec{T}/3$ & $\sqrt{\frac{10}{11}}\BB_{0,1}^\oslash  + {\curl}\frac{1}{16} \sqrt{\frac{5}{286}} r^2 \left(5 r^2-7\right) (3 \cos (2 \theta )+1)\vec{r}$\\
     $\BB_{0,3}^\oslash$ & $\BB_{101}^\vec{S},\BB_{111}^\vec{S}/3,\BB_{211}^\vec{S}/3$ & $\sqrt{\frac{10}{11}}\BB_{0,1}^\oslash + {\curl}{\curl} \frac{1}{32} \sqrt{\frac{5}{286}}r^2 \left(5 r^2-7\right) \cos (\phi ) \sin (2 \theta )\vec{r}$\\
     $\BB_{0,4}^\oslash$ & $\BB_{101}^\vec{S},\BB_{111}^\vec{S}/3,\BB_{211}^\vec{T}/3$ & $\sqrt{\frac{10}{11}}\BB_{0,1}^\oslash + {\curl}\frac{1}{8} \sqrt{105} r^2 \left(r^2-1\right) \cos (\phi ) \sin (2 \theta )\vec{r}$\\
     $\BB_{0,5}^{\oslash,\dagger}$ & $\BB_{101}^\vec{S},\BB_{111}^\vec{S}/3,\BB_{101}^\vec{T}/3$ & $\sqrt{\frac{10}{11}}\BB_{0,1}^\oslash + {\curl}\frac{3}{4} \sqrt{\frac{35}{11}} f^t_{101}\vec{r}$\\
     $\BB_{0,6}^{\oslash,\dagger}$ & $\BB_{101}^\vec{S},\BB_{111}^\vec{S}/3,\BB_{201}^\vec{S}/3$ & $\sqrt{\frac{10}{11}}\BB_{0,1}^\oslash + {\curl}{\curl} \frac{1}{16} \sqrt{\frac{5}{286}}f^s_{201} \vec{r}$\\
     
	\end{tabular}
 
 with $f^s_{101}= r \left(3 r^2-5\right) \cos (\theta ) $,  $f^t_{101}= r \left(r^2-1\right) \cos (\theta )$, $f^s_{201} = r^2\left(5 r^2-7\right) (3 \cos (2 \theta )+1)$, $f^s_{111} = r \left(3 r^2-5\right) \cos (\phi ) \sin (\theta )$. $^\dagger$ Results for these background magnetic fields are shown in the supplementary material.
	\vspace*{-4pt}
\end{table}

We compute dense mode spectra of the considered background magnetic fields, as described in Section \ref{subsec:eigen}.
For the axisymmetric fields, a resolution of $N=80$ is considered, for each $m \in [0,17]$. 
We only consider $m\leq 17$, as the modes at larger $m$ are, by definition, not going to be observationally relevant.
Non-axisymmetric fields require a lower resolution for a given matrix size; here we find dense solutions using $N=40$. 
At this resolution the matrix dimensions are challenging $\mathcal{S} \sim 4.5\cdot 10^4$, requiring significant memory ($\sim 150$ GB) and computational effort for the full dense spectrum.

\begin{figure}
    \begin{minipage}{0.54\textwidth}
    (a)
    
    \includegraphics[width=\linewidth]{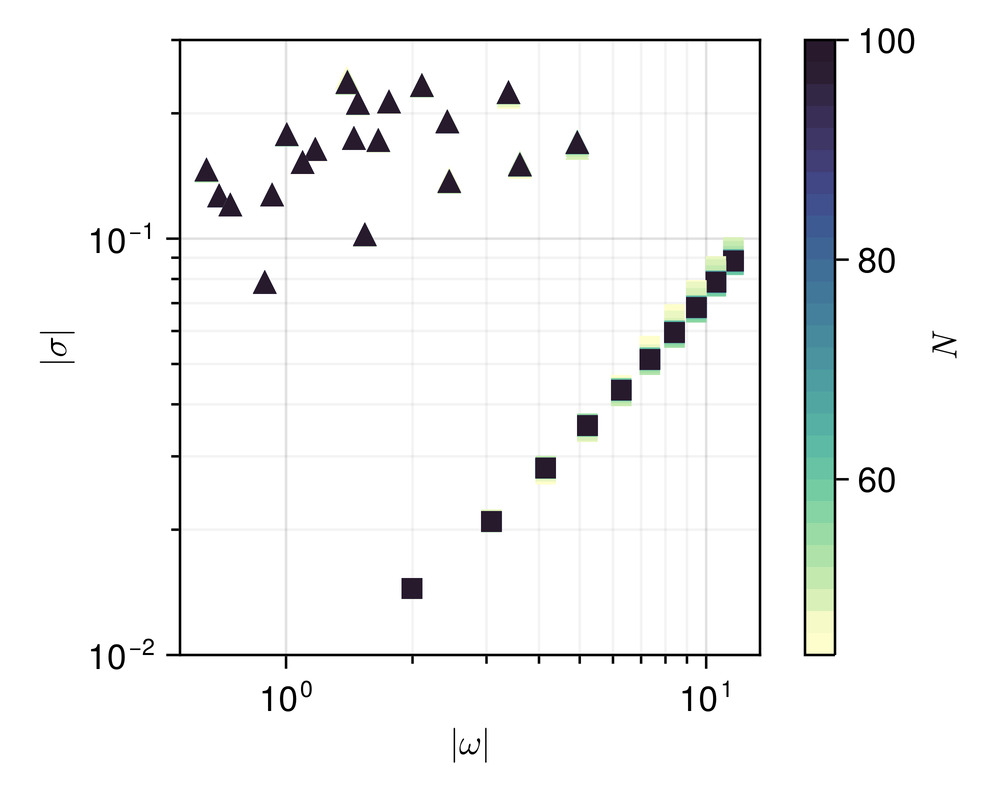}
    \end{minipage}
    \begin{minipage}{0.45\textwidth}
    (b)

    \includegraphics[width=\textwidth]{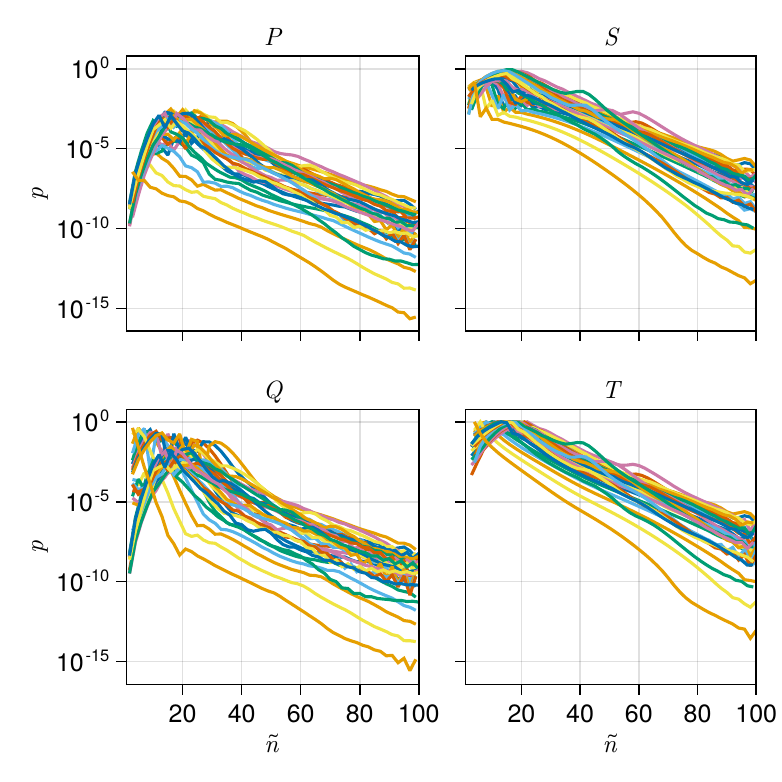}
    \end{minipage}
    
    \caption{Convergence of the observationally relevant modes for the background magnetic field $\BB_{0,1}^\oslash$. (a) Frequency-damping rate spectrum coloured by the truncation degree $N \in [44,100]$. (b) The energy density $p$ as a function of Cartesian polynomial degree $\tilde{n}$ for the poloidal ($P$ and $S$) and toroidal scalar ($Q$ and $T$) of the velocity and magnetic field (respectively) at the truncation $N=100$. Each colour corresponds to one mode.}
    \label{fig:convergence_S101S111}
\end{figure}

We confirm the convergence of the modes for the non-axisymmetric $\BB_0^\oslash$, which have been calculated at a truncation degree $N=40$, by tracking the observationally relevant modes up to $N=100$.
The frequency-damping rate spectrum at each truncation degree is shown in Figure \ref{fig:convergence_S101S111}(a), showing that the frequencies are almost unchanged and only the damping rates alter as the solution converges.
The spectra of the eigenvectors decays to less than $10^{-5}$ of the peak energy density at the truncation degree (shown in Figure \ref{fig:convergence_S101S111}(b)), giving us confidence in the spatial structure of all the observationally relevant modes that are further analysed.

\subsection{Mode spectra}

\begin{figure}
    \centering

    \includegraphics[width=0.95\textwidth]{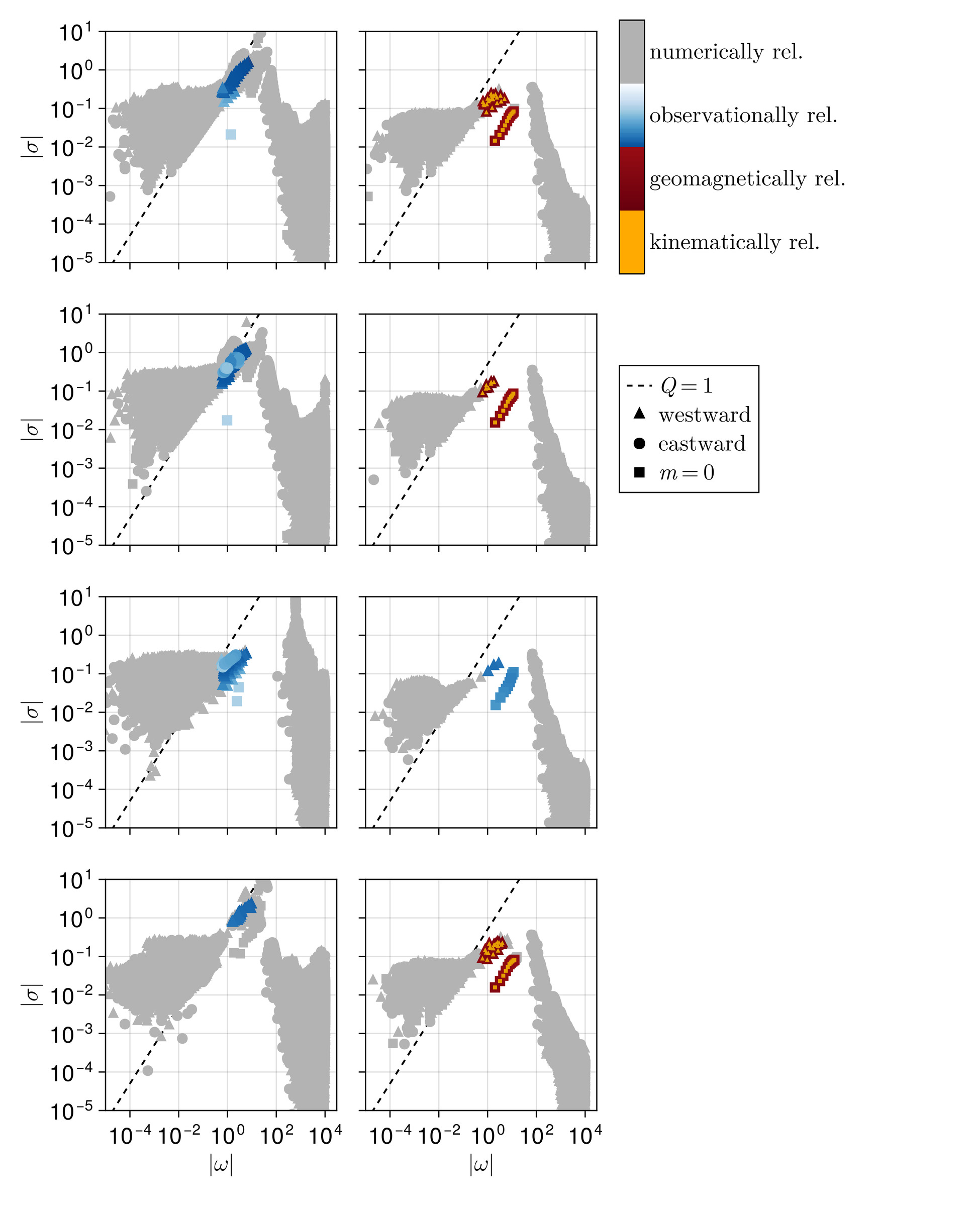}
    \caption{Frequency-damping rate spectra for the background magnetic fields in Table \ref{table:b0new}. The left and right column correspond to axisymmetric $\BB^\odot_{0,1-4}$ and non-axisymmetric $\BB^\oslash_{0,1-4}$, respectively. Dashed lines corresponds to $Q = |\omega/2\sigma| = 1$. Rectangles indicate modes with a dominant (or exact) azimuthal wave number $m=0$ in the velocity. Triangles and circles indicate a westward and eastward phase velocity of the mode, respectively.}
    \label{fig:spectra}
\end{figure}

The dense mode spectra, showing the frequency against the damping rate, are presented in Figure \ref{fig:spectra}.
For all considered $\BB_0$ we find numerically relevant modes throughout a broad frequency range (shown in grey).
Some numerically relevant modes in the observationally relevant frequency window satisfy also the other constraints (highlighted in the respective colours), and these modes are discussed in more detail.
We also report the direction of wave propagation: westward means that $\omega/m>0$ in our convention. 
When all azimuthal wave numbers are coupled, $m$ is determined as the azimuthal wave number of peak energy in $Q$.

Outside of that window, many inertial modes are clustered at high frequencies up to twice the diurnal frequency ($=2/\Le$, in terms of the Alfv\'en frequency). 
Their damping rates are small compared with their frequency, i.e. they have quality factors $Q \gg 1$. Magnetic diffusion (which is the only diffusive term present) is therefore negligible, consistent with the  irrelevance of the magnetic field in the dynamics of these modes.
It is noteworthy that some of the inertial modes approach the observationally relevant frequency range.
At frequencies below the observationally relevant frequency window we find MC modes. 
Their periods are mostly gathered around $\omega \sim \Le$, but extend to even lower periods, as well as up to the observationally relevant frequencies ($\omega\sim 1-10$).
At the Lundquist number considered here ($2\cdot 10^4$), most MC modes have a quality factor smaller than 1, i.e. they are over-damped. 
The number of numerically relevant MC modes with $Q>1$ at relevant periods depends on the background magnetic field.
For example, for $\BB_{0,3}^\odot$ (shown in the Figure \ref{fig:spectra}, third from the top, left column), there exist more MC modes at frequencies $\omega\sim 10^{-1}$ with large quality factors compared with other background magnetic fields.
For $\BB_{0,4}^\odot$, some of the MC modes at long periods are actually unstable (Figure \ref{fig:spectra}, bottom left), travelling eastward.
We are uncertain of the origin of this instability, but we note that for this background magnetic field the amplitudes are stronger in the deep interior of the domain.
Due to the stronger shear in the background magnetic field, conditions for a field gradient instability may be satisfied \cite{acheson_local_1983, fearn_core_2007}.

Observationally relevant modes (as defined in Section \ref{sec:data}), are highlighted in shades of blue (the different shades correspond to geomagnetic relevance; for our purposes here each shade of blue means observationally relevant).  
For each $\BB_0$ considered, except the axisymmetric purely toroidal field $\BB_{0,2}^\odot$ (shown in the suppl. material), observationally relevant modes are present.
The number of these modes is not the same between the different $\BB_0$. In addition, differences in rate of numerical convergence for each case, and the different resolutions for axisymmetric and non-axisymmetric calculations, making a direct comparison difficult.

Among the observationally relevant modes are torsional modes, mostly distinct from the other modes by their larger quality factor and a dominant $m=0$ velocity structure (shown as rectangles in the spectra).
The fundamental torsional mode has a frequency around $\omega =1$, with only slight differences in the frequency between each $\BB_0$. 
For the axisymmetric fields considered, only the gravest few torsional modes are numerically relevant at this resolution ($N=80$), whilst for the non-axisymmetric case many torsional modes of higher degrees also are converged at a lower truncation of $N=40$.
We can understand this difference in numerical convergence from the properties of the reduced ideal 1D torsional mode equation, which is only solvable when (i) $v_A$, given by \eqref{eq:va}, does not vanish at the axis and (ii) if $v_A(s=1)=0$ then $v_A \sim (1-s)^\nu$ with $\nu\leq 1$ \cite{maffei_propagation_2016}.
Luo \& Jackson \cite{luo_waves_2022a} showed that torsional modes can exist for a $\BB_0$ that fail to satisfy these conditions, when magnetic diffusion is present. 
However, a high resolution is needed to resolve the thin structures arising near the axis and equator. 
This is why, at the resolution considered here, only the gravest few torsional modes are numerically relevant for the axisymmetric $\BB_0^\odot$.
For the non-axisymmetric $\BB_0^\oslash$ discussed here, $v_A(s)$ is non-zero everywhere, so no such pathological points exist. The solutions are therefore larger scale because signficant magnetic diffusion is not required on the axis or equator.  
With the exception of $\BB_{0,4}^\odot$ (bottom left in Figure \ref{fig:spectra}), all the gravest torsional modes (and, if present, most of the higher degree torsional modes) are observationally relevant.

Besides torsional modes, we find that some MC modes are also observationally relevant.
Their quality factors are on the order of 1--10, smaller than those of torsional modes, but observationally relevant. 
Their quality factors should also be larger for larger values of $\Lu$.
For all non-axisymmetric $\BB_0^\oslash$, as well as $\BB_{0,1}^\odot$ and $\BB_{0,4}^\odot$, all observationally relevant MC modes are westward propagating. For $\BB_{0,2}^\odot$ and $\BB_{0,3}^\odot$, some observationally relevant modes are also eastward propagating.

\subsection{Columnarity}
Columnarity of flows, in which the Coriolis force is approximately in balance with the pressure gradient, is believed to be important in the rapidly rotating dynamics of Earth's core, at time scales close to the Alfv\'en period \cite{jault_axial_2008}. 
We compute a measure of columnarity similar to \cite{luo_waves_2022b}, as
\begin{equation}
    \chi = 4\pi\int\left(\frac{\left<u_s\right>^2 + \left<u_\phi\right>^2}{\left<u_s^2+u_\phi^2\right>}\right)^{1/2}s\,\mathrm{ds},
\end{equation}
with $\left<.\right> = (4\pi H)^{-1} \oint\int . \, \mathrm{d}z\mathrm{d}\phi$.
The factor $4\pi$ ensures that $\chi=1$ when the flow is perfectly columnar.

\begin{figure}
    \centering
    \includegraphics[width=0.7\textwidth]{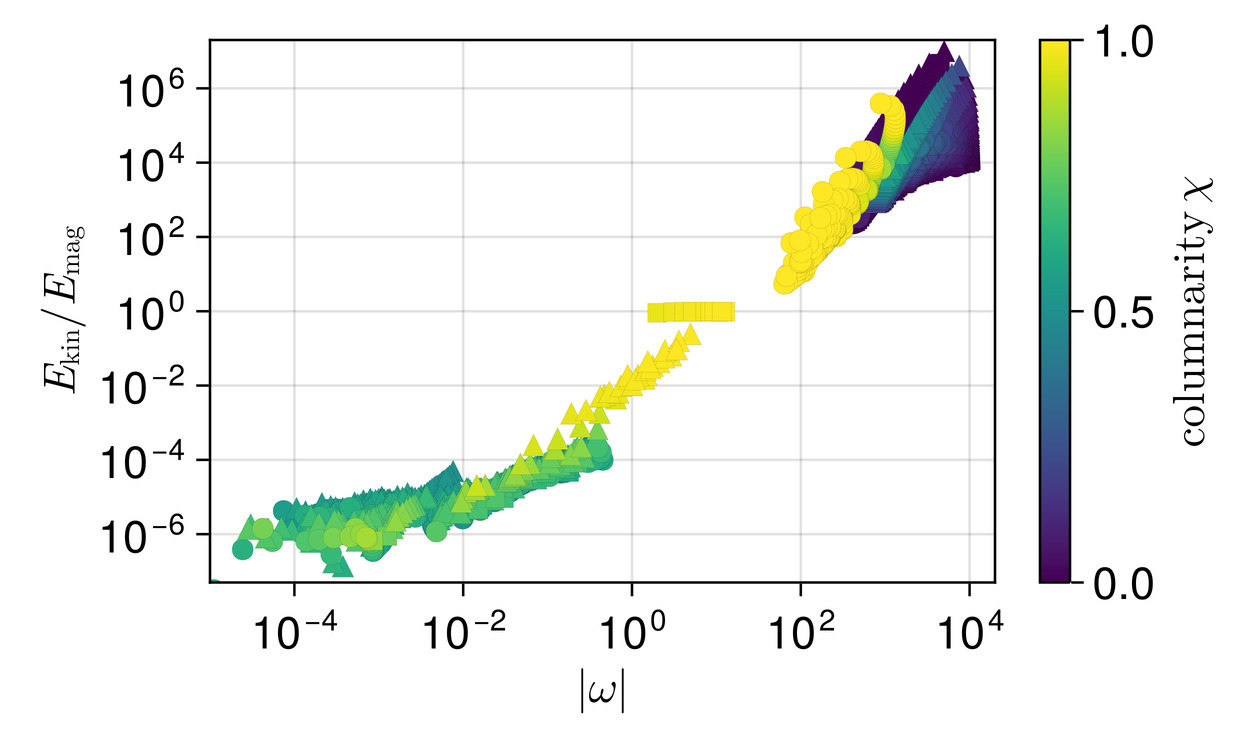}
    \caption{Ratio of kinetic to magnetic energy as a function of frequency $\omega$ for the mode spectrum calculated for the background magnetic field $\BB_{0,1}^\oslash$, at $N=40$. Shown are only the numerically relevant modes.}
    \label{fig:ekmspectrum}
\end{figure}

In Figure \ref{fig:ekmspectrum} we present the ratio of kinetic to magnetic energy as a function of frequency, coloured by the columnarity, for one non-axisymmetric magnetic field.
It is evident that modes closer to the Alfv\'en frequency are more columnar than modes that are much faster or slower.
Keeping in mind our restriction to numerically relevant modes here, smaller scale modes with low columnarity may fill the spectrum at the same frequencies, but they are not of importance from an observational perspective.
Despite having a much stronger magnetic energy compared with the kinetic energy, the westward propagating MC modes are all very columnar, i.e. quasi-geostrophic (QG), confirming the validity of previous studies that, a-priori, imposed quasi-geostrophy on the flow to investigate interannual QGMC modes \cite{gerick_fast_2021, gillet_satellite_2022}.
At the highest and lowest frequencies, non-columnar modes are found.
For these modes either the strong inertial force or Lorentz force, respectively, dominate the flow structure.
In a related axisymmetric case, columnar MC modes have have been presented in Luo et al. \cite{luo_waves_2022b}.
In their work, no columnar modes were found for the background magnetic field $\BB_{0,1}^\odot$ (which has only a $\BB^\vec{S}_{101}$ component) at the resolution they considered ($N\sim40$).
At the higher resolution considered here, we find a QGMC mode branch, showing that a relatively high resolution is needed in this particular background magnetic field to find adequate convergence of these modes.
A lack of an azimuthal component does not seem to be the relevant property of $\BB_0$ to observe these columnar modes.

\subsection{Geomagnetic relevance}

To put all observationally relevant modes into context of the magnetic field variation at the surface of the core, the region that we can access through observations on Earth, we calculate the weighted temporal and longitudinal rms of the radial magnetic field variation, $\mathrm{rms}_{t,\phi}\partial_t B_r \sin\theta$.
For the modes, the temporal rms is computed by taking the absolute value of the complex spatial magnetic field, reconstructed by \eqref{eq:mag_basis}. This produces the exact temporal rms in the limit of large $Q$, that we assume here.

\begin{figure}
    \centering
    
    \includegraphics[width=0.8\textwidth]{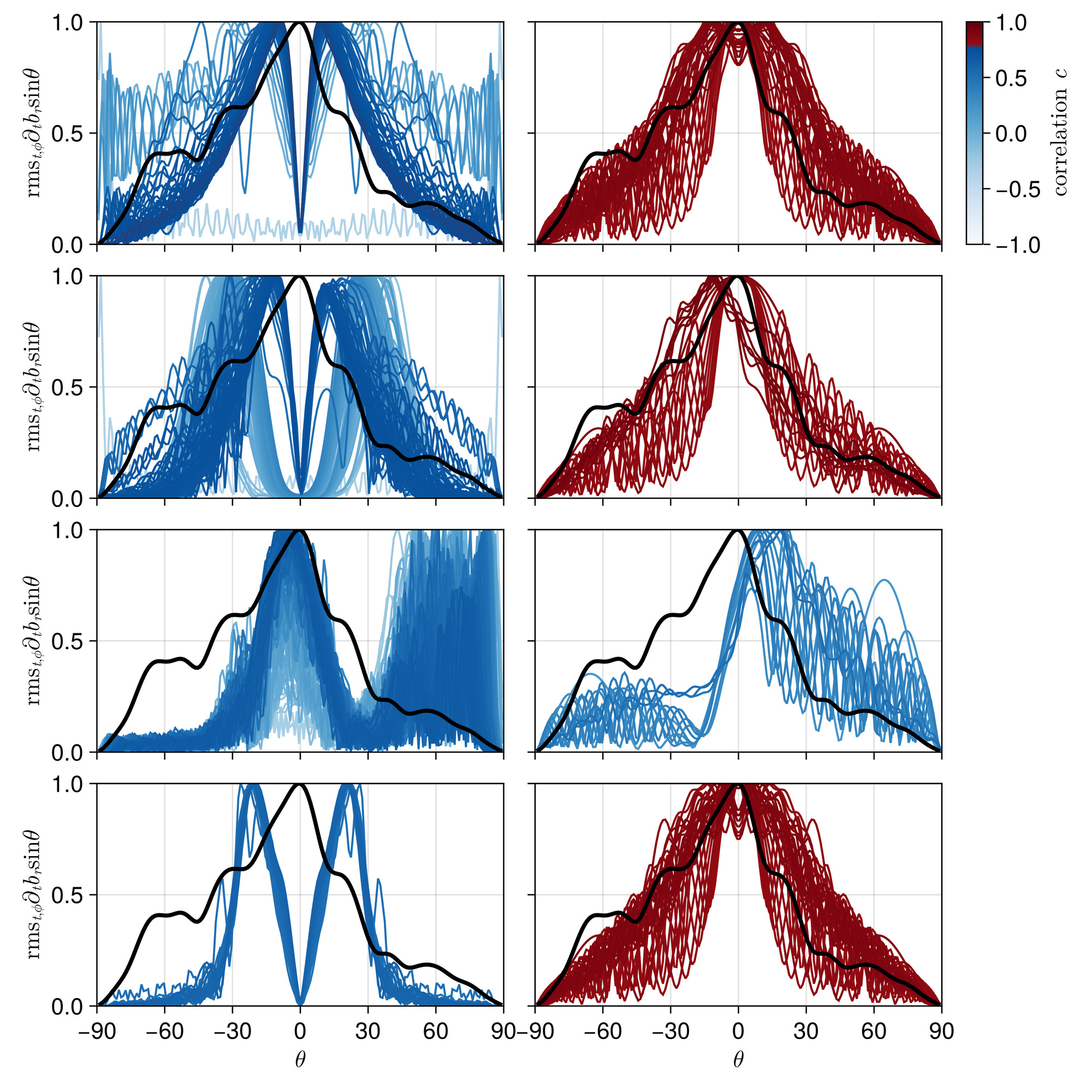}
    
    \caption{Weighted temporal and longitudinal rms of the radial magnetic field changes for each mode from background magnetic fields considered. The left and right column correspond to axisymmetric fields $\BB_{0,1-4}^\odot$ and non-axisymmetric fields $\BB_{0,1-4}^\oslash$, respectively. Each coloured lines corresponds to a single observationally relevant mode. The colour indicates the correlation to the same rms of the CHAOS-7.16 model, with blue colours for geomagnetically irrelevant modes ($c<0.8$) and red for geomagnetically relevant modes ($c>0.8$). The rms of the filtered (1--23.7\,yr) SV derived from CHAOS-7.16 is shown in thick black lines.}
    \label{fig:rms_thetas}
\end{figure}

A comparison is then made to the rms derived from the observations (CHAOS-7.16 model), as shown in Section \ref{sec:data}, to determine the geomagnetical relevance of the modes. 
In Figure \ref{fig:rms_thetas} we present the rms profiles of the modes compared with the rms of the observations, for each considered background magnetic field.
The colour of the profiles of the modes corresponds to the correlation $c$ with the observed rms.

When the correlation is larger than the threshold value that determines geomagnetic relevance ($0.8$), the colour is red instead of shaded blue ($c<0.8$).
It is found that none of the axisymmetric magnetic fields shows modes with geomagnetically relevant correlation.
This low correlation is mainly due to the vanishing radial rms component at the equator ($\theta = 0$) found for all modes, for background magnetic fields that have a vanishing radial component at the equator.
This suggests that any combination of modes for such magnetic fields is unable to reproduce the observed rms SV.
When $\BB_0 \cdot \vec{r} \neq 0$ at the equator for an axisymmetric field (cf. Figure \ref{fig:rms_thetas}, third from the top on the left, corresponding to $\BB_{0,3}^\odot$, which has a quadrupolar component $\vec{B}^\vec{S}_{201}$), we find a large rms near the poles, which is not observed on Earth.
This strong hemispheric asymmetry through the quadrupolar component is also present in the non-axisymmetric field $\BB_{0,3}^\oslash$, that includes a $\vec{B}^\vec{S}_{211}$ component (shown in Figure \ref{fig:rms_thetas}, third from the top on the right).
However, for $\BB_{0,3}^\oslash$ there is no peak in the amplitude at high latitudes.
Unlike in the axisymmetric case, the non-axisymmetric fields considered here all show a peak rms in $\partial_t B_r$ near the equator, and smaller rms at high latitudes.
For the background magnetic fields $\BB_{0,(1,2,4)}^\oslash$, a high correlation between the observed rms and the rms of the modes is found, deeming them geomagnetically relevant according to our definition.
This is true for all modes that are observationally relevant, both the QGMC modes and the torsional modes.

We find that all modes for the non-axisymmetric $\BB_{0,1,2,4}^\oslash$ that are geomagnetically relevant are also kinematically relevant (see Figure \ref{fig:spectra}, right column). 
By our definition, this means that the peak in azimuthal velocity is near the equator ($|\theta|<30^\circ$).
This is in agreement with previous calculations of these modes in 2D reduced QG models \cite{gerick_fast_2021, gillet_satellite_2022}.

\subsection{Spatial structures}

\begin{figure}
    \centering
    \begin{minipage}{0.49\textwidth}
    (a)  $c = 0.78, \lambda = -0.98 + 4.36\mathrm{i}$
    
    \includegraphics[width=\linewidth]{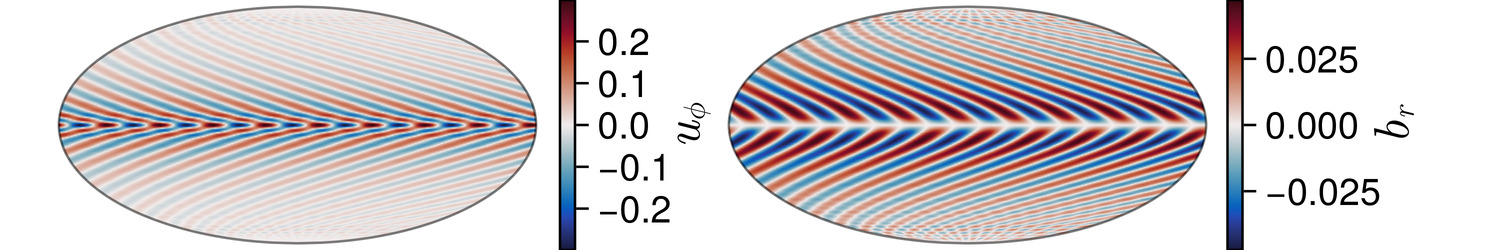} 

    (c) $c = 0.75, \lambda = -0.66 + 2.62\mathrm{i}$
    
    \includegraphics[width=\linewidth]{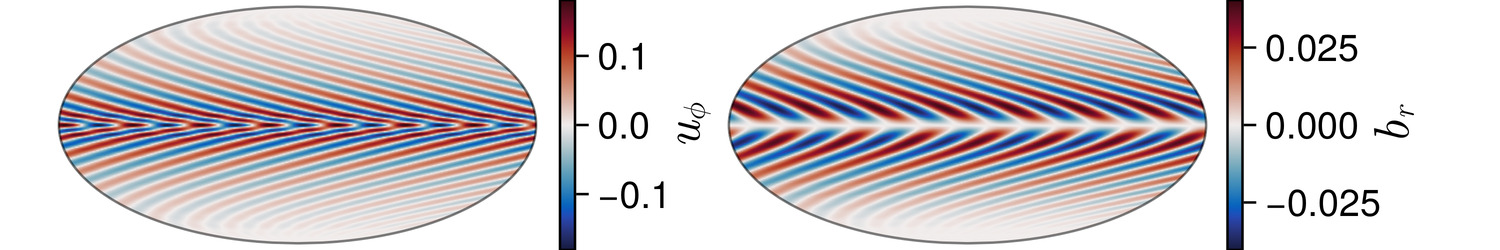} 

    (e) $0.68, \lambda = -0.21 + 1.23\mathrm{i}$
    
    \includegraphics[width=\linewidth]{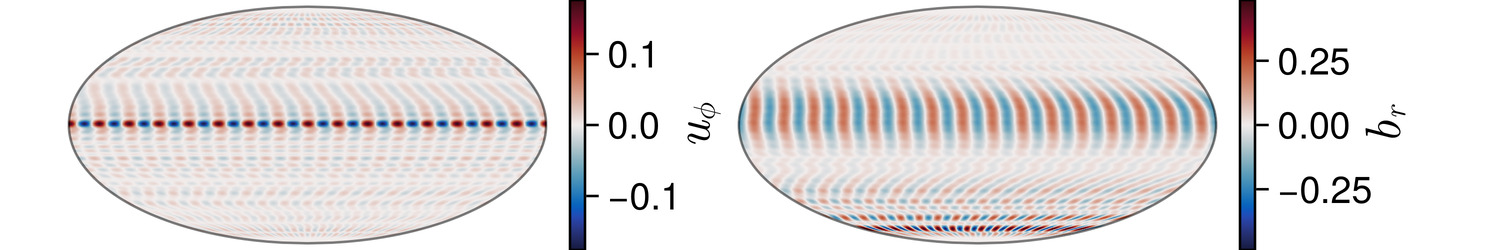} 

    (g) $c = 0.62, \lambda = -2.22 + 9.57\mathrm{i}$
    
    \includegraphics[width=\linewidth]{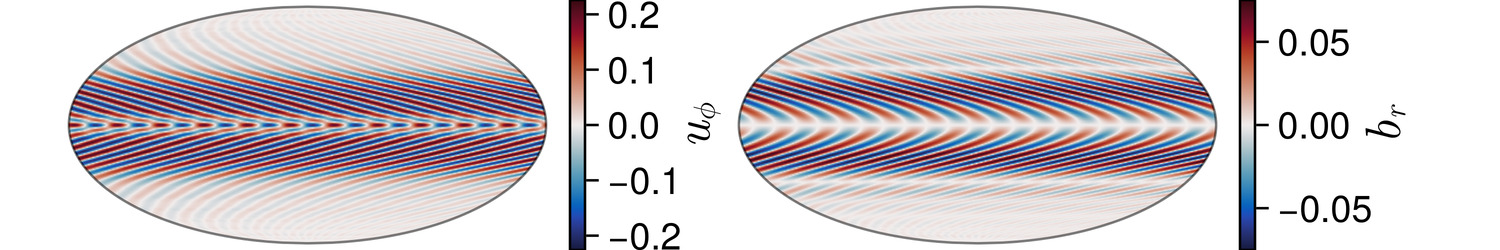} 
    \end{minipage}
    \begin{minipage}{0.49\textwidth}
    (b) $c = 0.91, \lambda = -0.22 + 3.38\mathrm{i}$
    
    \includegraphics[width=\linewidth]{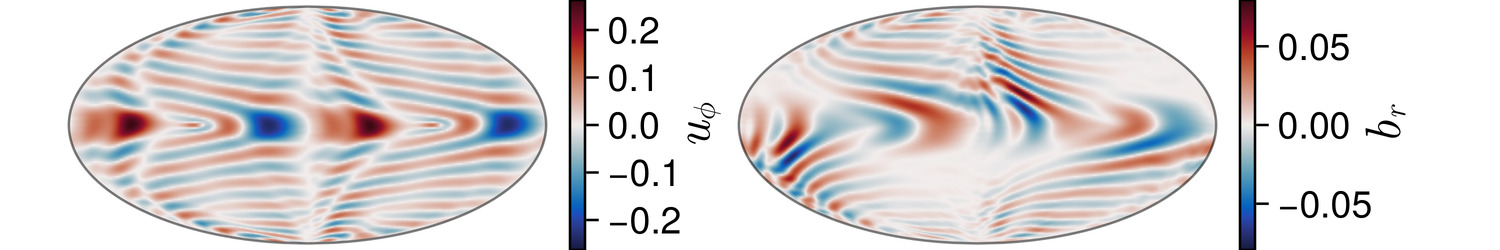} 

    (d) $c = 0.95, \lambda = -0.12 + 1.09\mathrm{i}$
    
    \includegraphics[width=\linewidth]{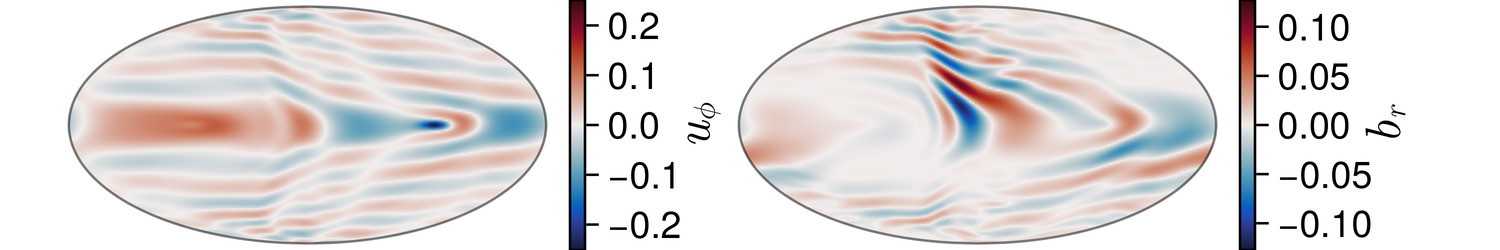} 

    (f) $c = 0.52, \lambda = -0.15 + 1.8\mathrm{i}$
    
    \includegraphics[width=\linewidth]{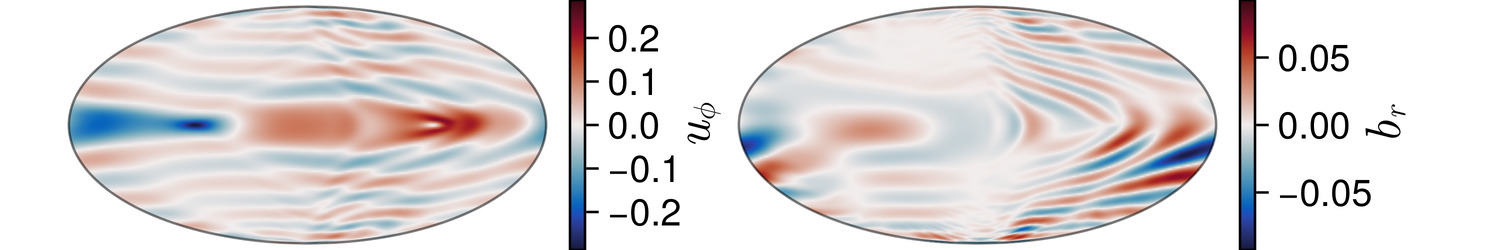} 

    (h) $c = 0.91, \lambda = -0.23 + 2.95\mathrm{i}$
    
    \includegraphics[width=\linewidth]{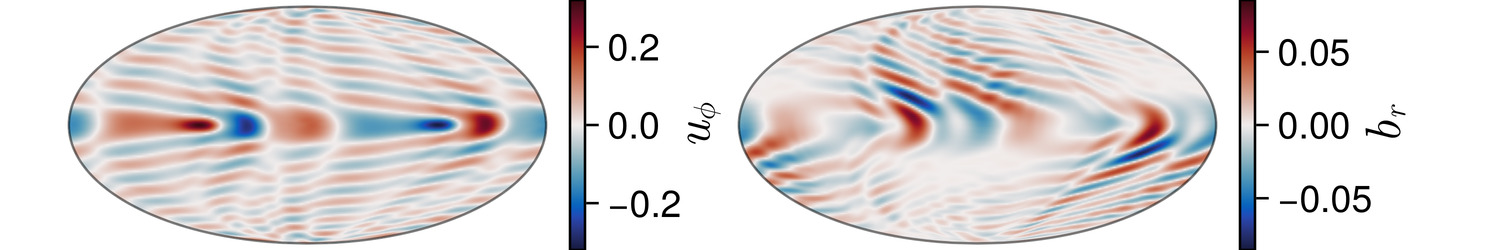} 
    \end{minipage}
    
    \caption{Azimuthal velocity (first column) and radial magnetic field (second column) for the mode of highest correlation in $\mathrm{rms}_{t,\phi} \partial_t B_r \sin\theta$ with the CHAOS-7.16 profile, for each considered background magnetic field. The modes for the axisymmetric fields $\BB_{0,1-4}^\odot$ (a,c,e,g) are truncated at $N=80$, and for the non-axisymmetric $\BB_{0,1-4}^\oslash$ (b,d,f,h) at $N=100$. Correlation $c$ and the eigenvalue are given for each mode. The units are Alfv\'en speeds, for both the magnetic field and the velocity, and amplitudes are arbitrary, so only their relative amplitude is relevant.}
    \label{fig:highest_corr_modes}
\end{figure}

For each background field, the spatial structure for the mode with highest correlation $c$ (see above) is shown in Figure \ref{fig:highest_corr_modes}. For each subpart (a-g), the left shows the azimuthal velocity and the right the radial magnetic field at the core surface.

A clear discrepancy between the modes for axisymmetric and non-axisymmetric $\BB_0$ can be seen. 
The modes of the axisymmetric fields (Figure \ref{fig:highest_corr_modes}a,c,e,g), all show a very large azimuthal wave number, both in the azimuthal velocity (left) and the radial magnetic field (right). 
There are observable modes with small $m$ for the axisymmetric fields as well, but they do not correspond to the mode of highest correlation.
The two highest correlating modes for $\BB_{0,1}^\odot$ and $\BB_{0,2}^\odot$ (shown in Figure \ref{fig:highest_corr_modes}a and c), are very similar in their spatial structure.
In both cases, the radial magnetic field vanishes at $\theta = 0$, whilst the amplitude is largest slightly above and below the equator.
The azimuthal flow is largest near the equator in both cases.
This similarity indicates that a toroidal field $\BB^\vec{T}_{101}$, which is added in $\BB_{0,2}^\odot$ on top of $\BB_{0,1}^\odot$, does not seem to affect strongly the structure of QGMC modes near the surface.
Of course, the toroidal field has some effect on the modes, recalling also the rms fields of all modes shown Figure in \ref{fig:rms_thetas} (first two plots on the top, left), and the two modes that are compared are not linked in any particular way for this comparison.
For $\BB_{0,4}^\odot$, the spatial structure is also similar to the modes of $\BB_{0,1}^\odot$ and $\BB_{0,2}^\odot$, but the amplitudes of the azimuthal velocity and the radial magnetic field are smaller at higher latitudes.
For the axisymmetric field $\BB_{0,3}^\odot$, the spatial structure at the surface is very small scale, with fine structures of highest amplitude of the azimuthal velocity near the equator and near the south pole for the radial magnetic field.
It is interesting to note that strong magnetic field perturbations can be spatially separated from velocity perturbations on the surface.
The small scale spatial structure of the modes in for axisymmetric $\BB_0^\odot$ is also evident in the slow spectral decay of the eigenvectors (shown in suppl. material Figure S2, and S3).

The highest correlating modes for the non-axisymmetric fields are very different compared with the axisymmetric ones.
The overall dominant spatial length scales are larger, and in all of the modes the velocity field shows a small azimuthal wave number combined with a larger cylindrical radial wave number.
The amplitude of both the azimuthal velocity and the radial magnetic field are largest near the equator for all non-axisymmetric $\BB_0^\oslash$ shown.
The biggest difference between the $\BB_0^\oslash$ shown is the longitudinal modulation of the amplitude of the radial magnetic field.

The low geomagnetic relevance for $\BB_{0,3}^\oslash$ is not very apparent from Figure \ref{fig:highest_corr_modes}(f) alone. 
However, the asymmetry about the equator in the peak amplitude of the radial magnetic field rms deems this background magnetic field configuration not geomagnetically relevant. 
The generally larger scale structure of the QGMC modes in the non-axisymmetric magnetic field configuration is highlighted also in the faster spectral decay of the eigenvectors compared with the modes of the axisymmetric fields (see suppl. material Figures S4-6).

\subsection{Dispersion of quasi-geostrophic Magneto-Coriolis modes}

We can investigate the dispersion of the QG-MC modes, i.e. the frequency as a function of cylindrical radial wave number $k_s$ and compare it with the dispersion relation
\begin{equation}
	\omega \approx -\frac{v_A^2 k_s^4 H^2}{2m\Omega},\label{eq:disp}
\end{equation}
derived by considering only the highest derivatives in $s$ in the reduced (QG) equations \cite[Appendix D]{gillet_satellite_2022}. 
Here, $v_A$ is the column averaged cylindrical radial Alfv\'en velocity
\begin{equation}\label{eq:v_A_qgmc}
    v_A(s,\phi) = \sqrt{\frac{1}{2H\mu_0\rho} \int_{-H}^H \left(\BB_0\bcdot\e_s\right)^2\,\mathrm{d}z}.
\end{equation}
This dispersion relation is only relevant when $k_s > m$ due to the neglect of derivatives in $\phi$, and when the background magnetic field is dominated by components that contribute to $V_A$. 
For example, if $\BB_0$ is an axisymmetric toroidal field $V_A$ vanishes and the dispersion relation cannot hold \cite{luo_waves_2022b}.

\begin{figure}
	\centering
    \begin{minipage}{0.49\textwidth}
    (a)
    
	\includegraphics[width=\linewidth]{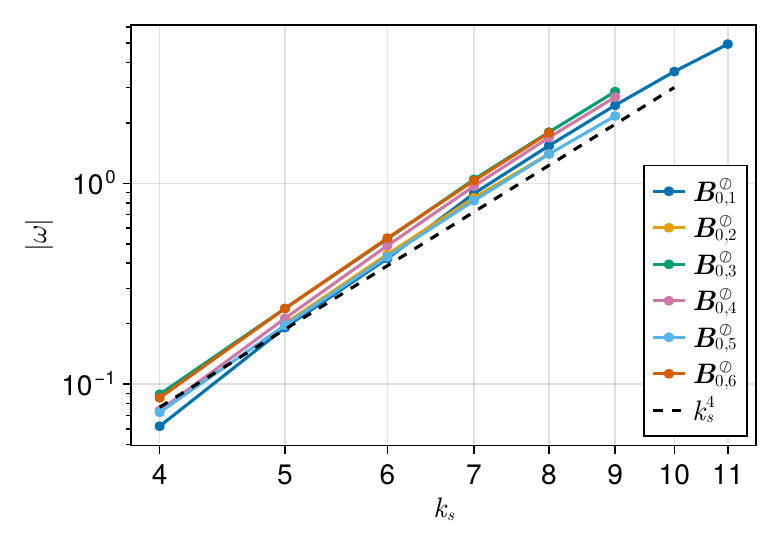}
    \end{minipage}
    \begin{minipage}{0.49\textwidth}
    (b)
    
    \includegraphics[width=\linewidth]{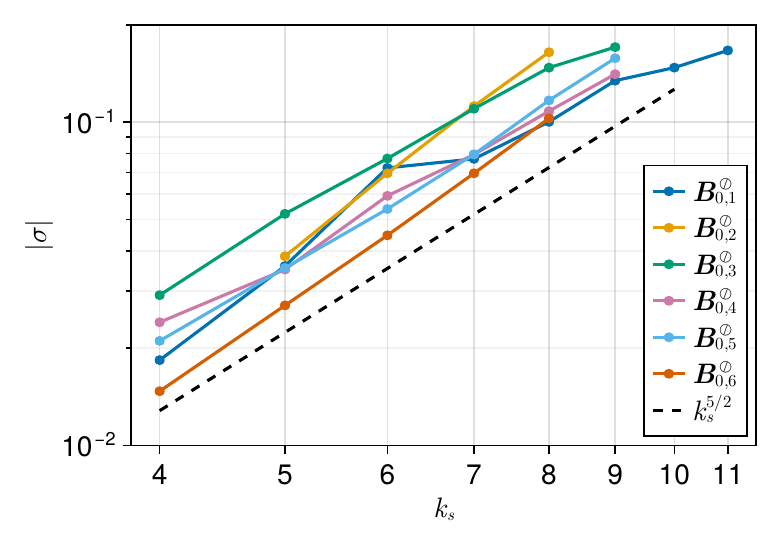}
    \end{minipage}
	\caption{Frequency (a) and damping rate (b) as a function of the cylindrical radial wave number $k_s$ for $m =1$ of the selected QGMC modes for each non-axisymmetric background magnetic field.}
	\label{fig:results_disp_relation}
\end{figure}

We select, for each non-axisymmetric field considered (see Table \ref{table:b0new}), several QGMC modes of dominant wave number $m=1$, without restricting the frequencies to the observationally relevant range.
Only the $m=1$ dominated QGMC modes are shown, as these are the modes that are reliably extracted from the dense spectra we can calculate at the computationally feasible resolution.
To ensure the eigenvalues are converged (on top of the numerically relevant constraint already imposed), we track the selected modes up to a higher resolution of $N=100$.
The cylindrical radial wave number $k_s$ is determined by counting alternating peaks in the azimuthal velocity.

Figure \ref{fig:results_disp_relation}(a) shows the dispersion relation for the non-axisymmetric background magnetic fields considered.
It is found that the spectra almost collapse, following closely a $k_s^4$ scaling, in agreement with \eqref{eq:disp}.
For large $m$ we expect this dispersion to be slightly different, as the derivatives along $\phi$ become more important and the $k_s^4$ scaling is likely no longer valid (at least for the moderate values of $k_s$ shown here).

We also show the damping rate as a function of $k_s$ for the same modes in Figure \ref{fig:results_disp_relation}(b).
The damping rates roughly follow a $k_s^{5/2}$ scaling, but strong variations to that scaling are observed between different background fields, which do not collapse in the same way as the frequencies do.

\section{Discussion}\label{sec:discussion}

We presented a suite of eigen mode calculations for several axisymmetric and non-axisymmetric background magnetic fields to investigate the sensitivity of modes in the interannual period range on the background magnetic field within the core.
Fully three-dimensional modes using non-axisymmetric magnetic fields that obey geophysically realistic boundary conditions have been calculated for the first time. 
The results underline the fact that non-axisymmetric magnetic fields are key to be able to produce geomagnetically relevant solutions in the interannual period range, relevant for global satellite based observations of the geomagnetic field.
The absence of observationally relevant modes for the purely axisymmetric toroidal field $\BB_{0,2}^\odot$ highlights the pathological character of such a simplified $\BB_0$. 
When comparing to rms fields derived from the CHAOS-7.16 (and CM6) model, the modes for the considered axisymmetric $\BB_0^\odot$ agree less than those for the non-axisymmetric $\BB_0^\oslash$.
The calculated $m=1$ dominated modes for $\BB_0^\oslash$ show good agreement with the dispersion relation \eqref{eq:disp}.

\begin{figure}

    \begin{minipage}{0.49\textwidth}
    (a)
    
      \includegraphics[width=\linewidth]{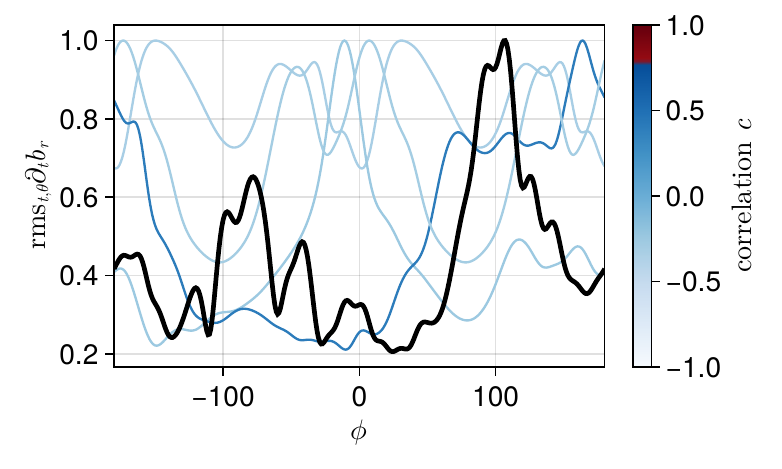}

    \end{minipage}
    \begin{minipage}{0.49\textwidth}
    (b)
    
    \includegraphics[width=\linewidth]{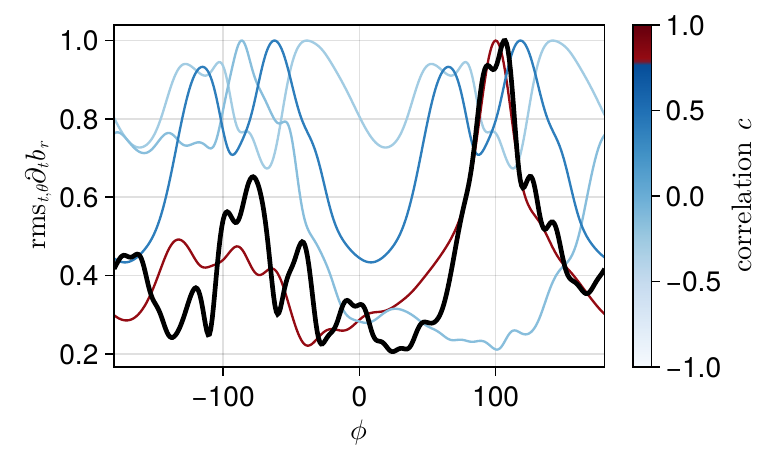}  
    \end{minipage}
    
    \caption{Temporal and latitudinal rms of the radial magnetic field changes for each mode of highest correlation (in $\mathrm{rms}_{t,\phi} \partial_t b_r \sin\theta$) from background magnetic fields $\BB_{0,1-4}^\oslash$. The colour indicates correlation in $\mathrm{rms}_{t,\theta} \partial_t b_r$ to the CHAOS-7.16 model (shown in thick black lines), with blue colours for $c<0.8$ and red colours for $c>0.8$. Panel (b) is identical to panel (a), but the mode profiles are shifted by $110^\circ$.}
    \label{fig:discussion_rms_long}
\end{figure}

There are likely more observationally relevant modes in the dense mode spectra, which are not captured at the numerical resolution used here.
However, we do not expect these additional modes to have an entirely different magnetic field structure at the CMB, in comparison to those that are already captured. 
Likely, these additional modes have smaller length scales only, whilst having overall similar properties.
To be able to compute at even higher resolutions, an iterative method should be used to compute a subspace of eigensolutions, as otherwise the matrix size becomes infeasibly large for dense calculations.
For example, one could sweep through a set of targets in the observationally relevant period range, or use contour-integral methods \cite{polizzi_densitymatrixbased_2009}.

A more realistic model for Earth should include a conductive inner core. 
However, at the considered periods (interannual to decadal), the inner core might be almost locked in to the motion of the fluid, therefore not contributing much to the dynamics of the modes.
In addition, the most geomagnetically relevant modes investigated here have their peak in amplitude near the equator both in the flow and the magnetic field and therefore an inner core might only play a minor role in their dynamics.

We have chosen the rms of the radial magnetic field variations as a proxy for geomagnetic relevance as a first step.
The temporal and longitudinal rms is able to constrain a peak amplitude near the equator. 
In addition, we can compare the mode of highest correlation in the longitudinal direction to the observations through the rms averaged over the latitude.
This comparison is made in Figure \ref{fig:discussion_rms_long}(a) and (b), where in (b) we have simply shifted the mode solutions arbitrarily by $\phi=110^\circ$.
We find that very little correlation in latitudinal rms is found between the geomagnetic SV and even the most favourable mode.
Shifting the modes by $110^\circ$, corresponding to a change in the orientation of the non-axisymmetric components of the background magnetic field (whose orientation was not chosen on geophysical grounds), we find that one mode (for $\BB_{0,2}^\oslash$) has a geomagnetically relevant correlation of $c>0.8$ with the rms derived from the observations.
This small experiment demonstrates how the background magnetic field may be constrained more accurately by taking into account the full information from the available observations.

In addition to the comparison to the observation, we compared the radial magnetic field variation as a function of longitude of modes with $v_A(\phi)$ of the background magnetic fields.
Here, $v_A(\phi)$ is the cylindrical radial average Alfv\'en speed $v_A(\phi)=\int v_A(s,\phi) s \mathrm{d}s$ , assumed to be relevant for QGMC modes.
No simple relationship between the observable radial magnetic field variation and $v_A(\phi)$ averaged through the bulk of the core was found, requiring further investigations.
From the simplified dispersion relation \eqref{eq:disp}, we expect to gain additional knowledge in the longitudinal direction of $v_A(s,\phi)$, compared with the information on $v_A(s)$ obtained through torsional modes.
There also does not seem to be a simple spatial relationship between the background magnetic field and the magnetic field perturbation of the modes, as shown in Figure \ref{fig:highest_corr_modes}, requiring further investigation.

Comparing the averaged rms profiles is clearly only a very crude way of determining geomagnetic relevance of the modes and the imposed background magnetic field.
However, so far hydromagnetic modes within the Earth's core have only been identified in the flow fields inferred from the geomagnetic data, and not from the geomagnetic data directly. 
Our work is a first step towards a more direct identification of these waves in the geomagnetic data.
One possible way to investigate this is to impose a background magnetic field matching the Earth's magnetic field at the core-mantle boundary, with varying structure in the bulk (potentially guided by the mean state of geodynamo simulations \cite{aubert_state_2023}), in order to further move towards a geophysically realistic model of the core.
In the future, additional satellite data from the recently launched Macau Science Satellite 1 \cite{zhang_novel_2022,livermore2024geomagnetic} and the prospective ESA mission NanoMagSat \cite{hulot_nanomagsat_2020} will further improve the data quality especially near the equator, the region that is most relevant for the observation of interannual QGMC modes. Except for the better spatial coverage in the observations, there is no reason to constrain this work to the satellite era.
Longer period modes, constrained by historic or archeomagnetic observations, could contribute new insights.

\enlargethispage{20pt}

\ack{The authors would like to thank two anonymous reviewers for their constructive comments. FG has received funding from the European Research Council (ERC) GRACEFUL Synergy Grant No. 855677.
This project has been funded by ESA in the framework of EO Science for Society, through contract 4000127193/19/NL/IA (SWARM + 4D Deep Earth: Core). This work was performed using HPC resources from the CNES Computing Center.}

\appendix

\section{Velocity and magnetic field bases terms}

\subsection{Inviscid velocity basis}\label{app:velocitybasis}

Poloidal and toroidal scalars for an inviscid velocity basis are given by
\begin{align}
    P_{ln} & = \sqrt{\frac{5+2l+4n}{4l(l+1)(n+1)^2}} r^l P_n^{(1,l+1/2)}\left(2r^2-1\right), & 0\leq n \leq \lfloor (N-l+1)/2-1\rfloor, \\
    Q_{ln} & = \sqrt{\frac{3+2l+4n}{l(l+1)}} r^l P_n^{(0,l+1/2)}\left(2r^2-1\right),         & 0\leq n\leq \lfloor (N-l)/2 \rfloor.
\end{align}
Note, the range of the radial degree $n$ depends on the truncation degree $N$ and the spherical harmonic degree $l$.

For these scalar functions, the resulting basis vectors are orthonormal, i.e.
\begin{equation}
    \int_\mathcal{V} \vec{P}^*_{lmn}\bcdot \vec{P}_{l'm'n'}\,\mathrm{d}V = \int_\mathcal{V} \vec{Q}^*_{lmn}\bcdot \vec{Q}_{l'm'n'}\,\mathrm{d}V = \delta_{ll'}\delta_{mm'}\delta_{nn'}.
\end{equation}

The projections onto the Coriolis term can be calculated analytically as
\begin{align}
    \int_\mathcal{V} \vec{P}^*_{lmn}\bcdot\left(\e_z\times\vec{P}_{l'm'n'}\right)\,\mathrm{d}V = \int_\mathcal{V} \vec{Q}^*_{lmn}\bcdot\left(\e_z\times\vec{Q}_{l'm'n'}\right)\,\mathrm{d}V = \frac{\mathrm{i}m}{l(l+1)}\delta_{ll'}\delta_{mm'}\delta_{nn'},
\end{align}
and
\begin{align}
    \begin{split}
        &\int_\mathcal{V} \vec{P}^*_{lmn}\bcdot\left(\e_z\times\vec{Q}_{l'm'n'}\right)\,\mathrm{d}V = -\int_\mathcal{V} \vec{Q}^*_{l'm'n'}\bcdot\left(\e_z\times\vec{P}_{lmn}\right)\,\mathrm{d}V = \\
        &\delta_{mm'}\left( \sqrt{\frac{(l^2-1)(l-m)(l+m)}{l^2(4l^2-1)}}\delta_{l(l'+1)}\delta_{nn'} + \frac{l+2}{l+1}\sqrt{\frac{l(l-m+1)(l+m+1)}{(l+2)(2l+1)(2l+3)}}\delta_{l(l'-1)}\delta_{n(n'+1)}\right).
    \end{split}
\end{align}

\subsection{Insulating magnetic field basis}\label{app:insulatingmfbasis}

Poloidal and toroidal scalars satisfying the insulating boundary conditions \eqref{eq:insulatingBC} are given by
\begin{align}
    S_{ln} & = f_s r^l\sum_{k=0}^{2}(-1)^k(1+\delta_{k1})(2l+4n+2k-3)P_{n-k}^{(0,l+1/2)}\left(2r^2{-}1\right), & 1\leq n \leq \lfloor (N-l+1)/2\rfloor, \label{eq:Sln} \\
    T_{ln} & =  f_t r^l\left(P_{n}^{(0,l+1/2)}\left(2r^2{-}1\right) - P_{n-1}^{(0,l+1/2)}\left(2r^2{-}1\right)\right), & 1\leq n \leq \lfloor (N-l)/2 \rfloor.
\end{align}
The normalization factors
\begin{align}
    f_s & = \left(2l(l+1)(2l+4n-3)(2l+4n-1)(2l+4n+1)\right)^{-1/2},\\
    f_t & = \left(\frac{(2l+4n-1)(2l+4n+3)}{2l(l+1)(2l+4n+1)}\right)^{1/2},
\end{align}
ensure that
\begin{equation}
    \int_{\mathbb{R}^3}\vec{S}_{lmn}\cdot\vec{S}_{lmn}^*\,\mathrm{d}V = \int_{\mathbb{R}^3}\vec{T}_{lmn}\cdot\vec{T}_{lmn}^*\,\mathrm{d}V = 1.
\end{equation}
The basis is not orthogonal, but tridiagonal, so that
\begin{align}
    \int_{\mathbb{R}^3}\vec{T}_{lmn}\cdot\vec{T}_{l'm'n'}\,\mathrm{d}V & = \delta_{ll'}\delta_{mm'}\left(\delta_{nn'}-\frac{1}{2}\sqrt{1 - \frac{3}{2l + 4(n-1) + 1} + \frac{3}{2l + 4(n-1)+5}}\delta_{(n-1)n'}\right) \\
    \int_{\mathbb{R}^3}\vec{S}_{lmn}\cdot\vec{S}_{l'm'n'}\,\mathrm{d}V & = \delta_{ll'}\delta_{mm'}\left(\delta_{nn'}-\frac{1}{2}\sqrt{1 + \frac{3}{2l + 4n - 1} - \frac{3}{2l + 4n - 5}}\delta_{(n-1)n'}\right)
\end{align}

The basis is however orthogonal with respect to the vector Laplacian \cite{chen_optimal_2018a}
\begin{align}
	\begin{split}
    \int_{\mathbb{R}^3}\B_{lmn}\cdot\bnabla^2\B_{l'm'n'}\,\mathrm{d}V &= \int_\mathcal{V}\curl\B_{lmn}\cdot\curl\B_{l'm'n'}\,\mathrm{d}V\\
	&= -\frac{1}{2}\delta_{ll'}\delta_{mm'}\delta_{nn'}\begin{cases}
        (2l + 4n + 1)(2l + 4n - 3) \quad , \B = \vec{S} \\
        (2l + 4n - 1)(2l + 4n + 3) \quad , \B = \vec{T}.
    \end{cases}
	\end{split}
\end{align}

The magnetic field basis, designed to be orthogonal w.r.t the vector Laplacian, was found to behave asymptotically like the single Jacobi polynomial only when projecting over all space.
The coefficients of the poloidal scalar \eqref{eq:Sln} take a ratio of $[1,-2,1]$ for large values of $n$, reducing to a single Jacobi polynomial through the recurrence identities \cite[eq. (3)]{livermore_galerkin_2010}.
The same holds for the toroidal scalar, with coefficients $[1,-1]$.

\section{Projections of Lorentz force and induction term}\label{app:li_projections}

To calculate the projection of the Lorentz force, the individual projections are
\begin{subequations}
    \begin{align}
         & \begin{alignedat}{2}
            & \int_\mathcal{V} \vec{P}_i^* \bcdot \left(\left(\curl\vec{S}_j\right)\times\vec{S}_k\right)  \,\mathrm{d}V = \\
         & -\frac{A_{jki^*}}{2}\int P_{i} \left( \ell_i\left(\ell_j{+}\ell_k{-}\ell_i\right)(D_jS_j)\partial_r(r S_k) + \ell_k(\ell_j{-}\ell_k{+}\ell_i) r \partial_r ((D_jS_j) S_k) \right) \,\mathrm{d}r,
         \end{alignedat}\\
         & \begin{alignedat}{2}
               &\int_\mathcal{V} \vec{P}_i^* \bcdot \left(\left(\curl\vec{S}_j\right)\times\vec{T}_k + \left(\curl\vec{T}_k\right)\times\vec{S}_j\right)  \,\mathrm{d}V =\\
               &-E_{jki^*}\int P_{i}\left( \ell_i r(D_jS_j)T_k + (\ell_j{+}\ell_k{+}\ell_i)S_jT_k r^{-1}\right.\\
               &\left.-(\ell_j{+}\ell_k{-}\ell_i)\left(\partial_r(S_j T_k)+r\partial_r S_j\partial_r T_k\right) - \ell_krT_k\partial_r^2S_j-\ell_jrS_j\partial_r^2T_k \right) \,\mathrm{d}r,
           \end{alignedat} \\
         & \begin{alignedat}{2}
             & \int_\mathcal{V} \vec{P}_i^* \bcdot \left(\left(\curl\vec{T}_j\right)\times\vec{T}_k\right)  \,\mathrm{d}V = \\
             & - \frac{A_{jki^*}}{2} \int P_{i}\left( \ell_i\left(\ell_j{+}\ell_k{-}\ell_i\right)\partial_r(rT_j)T_k + \ell_j\left(\ell_k{+}\ell_i{-}\ell_j\right)r\partial_r(T_jT_k)  \right)\,\mathrm{d}r,
         \end{alignedat} \\
         & \int_\mathcal{V} \vec{Q}_i^* \bcdot \left(\left(\curl\vec{S}_j\right)\times\vec{S}_k\right)  \,\mathrm{d}V = - E_{jki^*}\int Q_{i}\ell_k\left(D_j S_j\right) S_k r  \,\mathrm{d}r,                                                                                                                                      \\
         & \begin{alignedat}{2}
             &\int_\mathcal{V} \vec{Q}_i^* \bcdot \left(\left(\curl\vec{S}_j\right)\times\vec{T}_k + \left(\curl\vec{T}_k\right)\times\vec{S}_j\right)  \,\mathrm{d}V = \\
             &\frac{A_{jki^*}}{2}\int Q_{i}\ell_k\left(\ell_k{-}\ell_j{-}\ell_i\right)\partial_r(rS_j)T_k-\ell_j\left(\ell_j{-}\ell_k{-}\ell_i\right)S_j\partial_r(rT_k)  \,\mathrm{d}r, \\
         & \int_\mathcal{V} \vec{Q}_i^* \bcdot \left(\left(\curl\vec{T}_j\right)\times\vec{T}_k\right)  \,\mathrm{d}V = E_{jki^*}\int Q_{i}\ell_jT_jT_kr \,\mathrm{d}r,
         \end{alignedat}
    \end{align}
\end{subequations}
with $\ell_i = l_i(l_i+1)$ and $D_iS_i = \partial_r^2 S_i + 2/r \partial_r S_i - l_i(l_i+1)/r^2 S_i$.

For the induction term, the projections are
\begin{subequations}
    \begin{align}
         & \begin{alignedat}{2}
               &\int_{\mathbb{R}^3} \vec{S}_i^* \bcdot \curl\left(\vec{P}_j\times\vec{S}_k\right)  \,\mathrm{d}V = \\
               & \frac{A_{jki^*}}{2}\left(\int_0^1 \left[S_i, \left(-\ell_j\left(\ell_i{+}\ell_k{-}\ell_j\right)P_j\partial_r(rS_k) + \ell_k\left(\ell_i{+}\ell_j{-}\ell_k\right) S_k\partial_r(rP_j)\right) r^{-2}\right]r^2\,\mathrm{d}r\right.\\
               & \left. \vphantom{\int_0^1} + l_i \ell_k\left(\ell_i+\ell_j-\ell_k\right)\left(\partial_r(P_j)S_kS_i\right)_{r=1}\right),
           \end{alignedat}                              \\
         & \int_{\mathbb{R}^3} \vec{S}_i^* \bcdot \curl\left(\vec{P}_j\times\vec{T}_k\right)  \,\mathrm{d}V = E_{jki^*}\int \left[S_i, \ell_j P_j T_k r^{-1}\right] r^2\,\mathrm{d}r,                                                                                                                 \\
         & \int_{\mathbb{R}^3} \vec{S}_i^* \bcdot \curl\left(\vec{Q}_j\times\vec{S}_k\right)  \,\mathrm{d}V =  E_{jki^*} \left(\int \left[S_i, \ell_k Q_j S_k r^{-1}\right] r^2\,\mathrm{d}r + l_i \ell_k (Q_j S_k S_i)_{r=1}\right),                                                                 \\
         & \begin{alignedat}{2}
               &\int_{\mathbb{R}^3} \vec{T}_i^* \bcdot \curl\left(\vec{P}_j\times\vec{S}_k\right)  \,\mathrm{d}V = \\
               & E_{jki^*}\int T_i\left(\left(\ell_i{+}\ell_j{+}\ell_k\right) P_jS_k r^{-1} - \left(\ell_j{+}\ell_k{-}\ell_i\right) \left(\partial_r(P_jS_k)+r\partial_r(P_j)\partial_r(S_k)\right)\right.\\
               & \left. - \ell_j r P_j \partial_r^2 S_k - \ell_k r S_k \partial_r^2 P_j\right)\,\mathrm{d}r,
           \end{alignedat}                                                   \\
         & \begin{alignedat}{2}
             &\int_{\mathbb{R}^3} \vec{T}_i^* \bcdot \curl\left(\vec{P}_j\times\vec{T}_k\right)  \,\mathrm{d}V = \\
             &\frac{A_{jki^*}}{2} \int T_i\left(-\ell_i\left(\ell_j{+}\ell_k{-}\ell_i\right) T_k\partial_r(rP_j)+\ell_j\left(\ell_j{-}\ell_i{-}\ell_k\right)r\partial_r(P_jT_k)\right) \,\mathrm{d}r, 
         \end{alignedat}\\
         & \begin{alignedat}{2}
             &\int_{\mathbb{R}^3} \vec{T}_i^* \bcdot \curl\left(\vec{Q}_j\times\vec{S}_k\right)  \,\mathrm{d}V =  \\
             &\frac{A_{jki^*}}{2} \int T_i\left(\ell_i\left(\ell_j{+}\ell_k{-}\ell_i\right) Q_j\partial_r(rS_k)-\ell_k\left(\ell_k{-}\ell_i{-}\ell_j\right)r\partial_r(Q_jS_k)\right) \,\mathrm{d}r, 
         \end{alignedat}\\
         & \int_{\mathbb{R}^3} \vec{T}_i^* \bcdot \curl\left(\vec{Q}_j\times\vec{T}_k\right)  \,\mathrm{d}V = E_{jki^*}\int T_i Q_j T_k r\, \mathrm{d}r,
    \end{align}
\end{subequations}
where $\left[S_i,F\right] = S_iF\ell_i+\partial_r(rS_i)\partial_r(rF)/r^2$.

\section{External contributions to the induction equation}
\label{app:external}

\subsection{Inner product}

Consider the poloidal magnetic field $\vec{S}_{lmn}$ in the interior and the associated exterior poloidal field $\hat{\vec{S}}_{lmn}$. The inner product over all space can be divided
\begin{align}
    \int_{\mathbb{R}^3} \vec{S}\bcdot\vec{S}'\,\mathrm{d}V = \int_V \vec{S}\bcdot\vec{S}'\,\mathrm{d}V + \int_{\hat V} \hat{\vec{S}}\bcdot\hat{\vec{S}}'\,\mathrm{d}V.
\end{align}
Focusing on the integral over the exterior domain, we find
\begin{align}
    \int_{\hat V} \hat{\vec{S}}\bcdot\hat{\vec{S}}'\,\mathrm{d}V & = ll'S_{ln}(1)S_{l'n'}(1)\int_1^\infty\int_0^\pi\oint \nabla I_m^l\bcdot\nabla I_{m'}^{l'}  r^2\sin\theta\, \mathrm{d}\phi\mathrm{d}\theta\mathrm{d}r \\
                                                  & = S_{ln}(1)S_{l'n'}(1) l^2(l+1) \delta_{ll'}\delta_{mm'},
\end{align}
where we used equation B.3.3 of \cite{livermore_magnetic_2004a}, but with fully normalised spherical harmonics.

\subsection{Induction term}

For an inviscid fluid, with $\vec{u}\neq \vec{0}$ at $r=1$, we need to calculate additional surface terms in the calculation of the magnetic induction term.

\begin{equation}
    \int_{\mathbb{R}^3} \B\cdot\curl\left(\vec{u}\times\B\right)\,\mathrm{d}V = \int_V \B\cdot\curl\left(\vec{u}\times\B\right)\,\mathrm{d}V + \int_{\hat V} \B\cdot\curl\left(\vec{u}\times\B\right)\,\mathrm{d}V
\end{equation}

Only poloidal magnetic fields are non-zero in the exterior, and we consider only
\begin{align}
    \int_{\hat V}\hat{\vec{S}}_i\cdot\curl\left(\vec{u}_j\times\hat{\vec{S}}_k\right)\,\mathrm{d}V & = l_i l_k S_i(1) S_k(1)\int_{\hat V}\nabla I_i\cdot\curl\left(\vec{u}_j\times\nabla I_k\right)\,\mathrm{d}V
\end{align}
where $\hat{\vec{S}}_i$ and $\hat{\vec{S}}_k$ are the respective potential fields associated with the poloidal components in the interior $\vec{S}_i$ and $\vec{S}_k$. We can further simplify
\begin{align}
    \int_{\hat V}\nabla I_i\cdot\curl\left(\vec{u}_j\times\nabla I_k\right)\,\mathrm{d}V & = \int_{\hat V}\bnabla\cdot \left( \left(\vec{u}_j \times \nabla I_k\right)\times \nabla I_i\right)\,\mathrm{d}V                                                  \\
    & = \int_{\partial \hat V} \left((\nabla I_i \cdot \vec{u}_j) \nabla I_k - (\nabla I_i \cdot \nabla I_k)\vec{u}_j\right) \cdot \vec{n} \, \mathrm{d}A \\
    & = \int_{\partial \hat V} (\nabla I_i \cdot \vec{u}_j) \left(\vec{n}\cdot\nabla I_k\right) \, \mathrm{d}A
\end{align}

The contribution from a toroidal velocity $\vec{Q}_j$ is given by
\begin{align}
    \int_{\partial \hat V} (\nabla I_i \cdot \vec{Q}_j) \left(\vec{n}\cdot\nabla I_k\right) \, \mathrm{d}A & = -\oint\int \left((\nabla I_i \cdot \curl \left(Q_j(r) Y_j\vec{r}\right)) \frac{\partial I_k}{\partial r}\right)_{r=1}\sin(\theta)\,\mathrm{d}\theta\mathrm{d}\phi \\
    & = (l_k + 1) Q_j(1) \oint\int Y_k \left(\frac{\partial Y_i}{\partial \theta}\frac{\partial Y_j}{\partial\phi} - \frac{\partial Y_i}{\partial\phi}\frac{\partial Y_j}{\partial \theta}\right) \,\mathrm{d}\theta\mathrm{d}\phi \\
    & = (l_k + 1) Q_j(1) E_{i j k}
\end{align}

In summary
\begin{equation}
    \int_{\hat V}\hat{\vec{S}}_i\cdot\curl\left(\vec{Q}_j\times\hat{\vec{S}}_k\right)\,\mathrm{d}V = l_i l_k(l_k + 1) Q_j(1) S_k(1) S_i(1) E_{ijk}
\end{equation}

The contribution from a poloidal velocity $\vec{P}_j$ is given by
\begin{align}
    \int_{\partial \hat V} (\nabla I_i \cdot \vec{P}_j) \left(\vec{n}\cdot\nabla I_k\right) \, \mathrm{d}A & = -\oint\int \left((\nabla I_i \cdot \curl\curl\left(P_j(r) Y_j\vec{r}\right)) \frac{\partial I_k}{\partial r}\right)_{r=1}\sin(\theta)\,\mathrm{d}\theta\mathrm{d}\phi \\
    & = (l_k + 1) \frac{\partial P_j}{\partial_r}\bigg|_{r=1} \oint\int Y_k \left(\frac{\partial Y_i}{\partial \theta}\frac{\partial Y_j}{\partial\theta} + \frac{1}{\sin^2(\theta)} \frac{\partial Y_i}{\partial\phi}\frac{\partial Y_j}{\partial \phi}\right) \sin(\theta)\,\mathrm{d}\theta\mathrm{d}\phi \\
    & = \frac{1}{2}(l_k + 1) \frac{\partial P_j}{\partial_r}\bigg|_{r=1}  \left(l_j(l_j+1)-l_k(l_k + 1)+l_i(l_i+1) \right) A_{i j k},
\end{align}
so that
\begin{equation}
    \int_{\hat V}\hat{\vec{S}}_i\cdot\curl\left(\vec{P}_j\times\hat{\vec{S}}_k\right)\,\mathrm{d}V = \frac{1}{2}l_i \ell_k\left(\ell_j-\ell_k+\ell_i \right)\frac{\partial P_j}{\partial_r}\bigg|_{r=1}  S_k(1) S_i(1) A_{ijk}.
\end{equation}


\bibliographystyle{RS}
\bibliography{cite}

\end{document}